\documentclass[aip, jcp, citeautoscript, amsmath, amssymb, reprint,
]{revtex4-2}
\usepackage[title]{appendix}
\usepackage{graphicx, tikz, orcidlink}
\usepackage{epstopdf}
\usepackage{ascmac}
\usepackage{ulem}
\usepackage{cancel}
\usepackage{here}
\usepackage{physics}
\usepackage{bm}
\usepackage{comment}

\newcommand*{\ddp}{{\frac{\partial}{\partial p}}}

\begin{document}


\title{Quantum hierarchical Fokker--Planck equations with U(1) gauge fields : Application to the Aharonov--Bohm ring}
\date{Last updated: \today}
\author{Hyeonseok YANG\orcidlink{0009-0007-1308-4724}}
\author{Shoki KOYANAGI \orcidlink{0000-0002-8607-1699}}
\author{Yoshitaka TANIMURA\orcidlink{0000-0002-7913-054X}}
\email[Author to whom correspondence should be addressed: ]{tanimura.yoshitaka.5w@kyoto-u.jp}
\affiliation{Department of Chemistry, Graduate School of Science,
Kyoto University, Kyoto 606-8502, Japan}

\begin{abstract}
We investigate a three-dimensional subsystem under a time-dependent U(1) gauge field coupled to rotationally invariant environments. To capture the dynamic behavior of the subsystem under thermal excitations and dissipations, it is imperative to treat the bath in a non-Markovian and nonperturbative manner. This is because quantum noise is constrained by the uncertainty principle, which dictates the relationship between the noise correlation time and the amplitude of the energy fluctuation.
To this end, we derive the hierarchical equations of motion (HEOM) incorporating the gauge field, enabling a rigorous investigation of the dynamics of the reduced subsystem.
Transforming the HEOM into the Wigner representation yields quantum hierarchical Fokker--Planck equations [U(1)-QHFPE] with U(1) gauge fields. These equations incorporate vector fields into the damping operators while preserving both gauge invariance and rotational symmetry.
To demonstrate the practical use of the formalism, the effects of a heat bath in the Aharonov--Bohm (AB) ring. Our investigation includes simulations of the equilibrium distribution, linear absorption spectra, and AB currents under thermal conditions. Within a rotationally invariant system–bath (RISB) model, we predict the emergence of a persistent current even in dissipative environments, provided the bath is non-Markovian and the temperature is sufficiently low. We also assessed the validity of the Caldeira--Leggett model in this context.
\end{abstract}


\maketitle

\section{Introduction}\label{sec:intro}
Macroscopic quantum phenomena, such as Bose-Einstein condensation\cite{BE_SFbook} and macroscopic tunneling,\cite{Caldeira1981,Caldeira1983,Leggett1984PhysRevB.30.1208,Chen1986,Leggett2001} which arise from quantum interference between macroscopic states, have been a fundamental subject in quantum physics. 
Decoherence and relaxation are the key to exploring quantum properties, as they determine the lifetime of quantum nature: When they are strong, the system behaves classically.\cite{T06JPSJ,T20JCP,T15JCP,TW91PRA,TW92JCP}  The study of decoherence in Superconducting Quantum Interference Devices (SQUIDs)\cite{Jaklevic1964,SQUIDClarke2004,SQUIDClarke2004Review} is motivated not only by its practical implications, such as advancing quantum superconductive technologies and nanoscale devices,\cite{ByersYang1961,1986HalperinRevModPhys.58.533,1991ii,Blasi_ACring2023,1986HalperinRevModPhys.58.533,1991ii,1992FukuyamaAndo,BookImry,1986ABPhysRevLett.56.386,1987Umbach10.1063/1.97887,levy1990magnetization,bluhm2009persistent,PC2010,Fomin2018}  but also by philosophical inquiries, such as the quantum measurement problem.\cite{Penrose1996,Leggett2001}

To explore the decoherence of SQUIDs within the framework of quantum mechanics, a harmonic heat bath is commonly employed. The electromagnetic effects are treated as U(1) gauge fields, consisting of three vector potentials and one scalar potential.
While extensive theoretical studies of SQUID systems have been conducted, most of these are restricted to equilibrium conditions.\cite{Caldeira1981,Caldeira1983,Leggett1984PhysRevB.30.1208,2009MatsumotoPhysRevLett.102.237003,2014NoriPhysRevB.89.224507,Cuozzo2024} 
This creates a significant gap in understanding, as dynamical analyses—critical for device design—remain underexplored, even through numerical simulations.

In addition, while the Caldeira--Leggett (CL) model\cite{Caldeira1981,Caldeira1983,Leggett1984PhysRevB.30.1208,Chen1986,Leggett2001} has been widely used to study macroscopic quantum effects; however, its validity has not been rigorously established. For example, although the SQUID system exhibits rotational symmetry, the bath model in the CL framework lacks this property. This inconsistency implies that the bath coordinate interacting with the system at an angle $\theta$ differs from the bath coordinate interacting at $\theta + 2\pi$.\cite{ST02JPSJ}  

The lack of rotational symmetry in the heat bath presents significant challenges in describing even simple phenomena, such as the rotational spectral bands of molecules. When the CL model is applied to study the damped free rotor, the rotational bands observed in linear absorption spectra—resulting from transitions between quantized angular momentum states—cannot be reproduced, regardless of the strength of the system-bath (S-B) interaction.\cite{ST02JPSJ,ST03JCP,IT18JCP} 
To overcome this limitation, it is necessary to introduce rotationally invariant system bath modes (RISB)
introduced by Gefen, Ben--Jacob, and Caldeira.\cite{1987RISBPhysRevB.36.2770} Thus the quantum dynamics in a two-dimensional (2D)\cite{IT18JCP}  and three-dimensional (3D) rotors\cite{IT19JCP,Lipeng3Drotar2019} have been explored.

This paper thoroughly examines the dynamics of a macroscopic quantum state subjected to electromagnetic fields within a dissipative environment. We consider a subsystem with the gauge fields interacting with 3D rotationally invariant baths. We then derive the reduced equations of motion for the subsystem by tracing over the bath degrees of freedom. 
Due to the constraints of quantum noise imposed by the uncertainty principle, which establishes a relationship between noise correlation time and energy fluctuation amplitude, the S-B interaction must be treated in a non-perturbative and non-Markovian framework, acknowledging the quantum entanglement between the system and the bath.\cite{T06JPSJ,T20JCP}

In this paper, we extend the hierarchical equations of motion (HEOM) formalism to develop the U(1)-HEOM, which maintains the U(1) symmetry of the gauge field, including its interaction with the heat bath. These equations are further expressed in phase space as quantum hierarchical Fokker--Planck equations [U(1)-QHFPE], making them well-suited for numerical simulations.

There is a growing interest in systems that exhibit quantum effects sensitive to gauge fields.\cite{FrancoNORI2019resolution,FrancoNORI2020gauge,FrancoNORI2021gauge,FrancoNORI2021QEDPhysRevResearch.3.023079} 
Studies of cavity-QED systems often rely on quantum optical master equations under Markovian and factorized assumptions.\cite{FrancoNORI2022gauge,FrancoNORI2023generalized} In contrast, the U(1)-HEOM and U(1)-QHFPE address these problems under realistic conditions, offering a rigorous framework to analyze gauge field effects on topological phase systems, including low-temperature thermal excitation and dissipation where quantum effects dominate. Potential applications include systems that exhibit the quantum Hall effect (QHE)\cite{thouless1982quantized,tong2016lecturesquantumhalleffect}, the Berry phase\cite{BerryPhase1984quantal}, and the Uhlmann phase\cite{uhlmann1986parallel, uhlmann1991gauge, uhlmann1995geometric}. This framework also enables discussions on thermal effects in Haldane systems\cite{HaldaneModel1988model}, Dirac materials, and Chern insulating systems\cite{MartinDelgado2013density, MartinDelgado2014two, MartinDelgado2014uhlmann, MartinDelgado2015symmetry}.

Here, we demonstrate the capabilities of our formalism by simulating the persistent current in the Aharonov--Bohm (AB) ring,\cite{
ByersYang1961,1986HalperinRevModPhys.58.533,1991ii,Blasi_ACring2023,1986HalperinRevModPhys.58.533,1991ii,1992FukuyamaAndo,BookImry,1986ABPhysRevLett.56.386,1987Umbach10.1063/1.97887,levy1990magnetization,bluhm2009persistent,PC2010,Fomin2018} 
as an example of the AB effect.\cite{1959Aaronov-Bohm,Tonomura1986PhysRevLett.56.792,2008NoriTonomura}  The AB rings, which are regarded as a superconductive device\cite{ByersYang1961,1986HalperinRevModPhys.58.533,1991ii,Blasi_ACring2023} or nano device,
\cite{1992FukuyamaAndo,BookImry,1986ABPhysRevLett.56.386,1987Umbach10.1063/1.97887,levy1990magnetization,bluhm2009persistent,PC2010} 
been studied in various ways, such as connecting multiple rings,\cite{1986ABPhysRevLett.56.386} embedding quantum dots,\cite{1995AB_dotPhysRevLett.74.4047,1995AB_dotPhysRevB.52.R14360,2002AB_dotPhysRevLett.88.256806,KATSUMOTO2003151,Ando2004PhysRevB.69.115307,Tanatar2005PhysRevB.71.125338,LI2009521,Hatano2011,Li2025} and increasing the number of terminals,\cite{2009ABMultiterminalPhysRevB.79.195443} and tunneling junctions.\cite{1993ABtunnelPhysRevB.47.2768,Hatano2011,Aharony2014} 
Although the fundamental properties of quantum effects in gauge fields have been well-established experimentally and theoretically across various systems, much remains to be explored regarding the dissipation and thermal effects induced by phonons, which influence coherence lifetimes.\cite{2015GreenDissipationLi} In metal and semiconductor nanorings, gate resistance is often a significant factor, whereas in molecular nanorings, the focus shifts to resistance within the conductor itself.\cite{2018GreenPhysRevB.98.125417,2020ODissipationdoi:10.1021/acs.jpcc.0c01706,2011ABdephPhysRevB.84.035323,2012ABdepha2dotsPhysRevB.86.115453,2016ABdephaLIU2016163}
Compared to phonon-based models, the RISB model provides greater flexibility in describing molecular rings, as it accounts for both rotational symmetry and anisotropy.\cite{AsymmetricAB2024,ABYiJinJIn_2018,2019CaoPhysRevB.99.075436} 

The structure of this paper is as follows.
Section \ref{sec:theory} outlines the U(1)-HEOM approach, developed from a 3D S-B model incorporating a gauge field. Utilizing the discretized Wigner transformation under periodic boundary conditions, the U(1)-QHFPE is derived. In Sec. \ref{sec:ABring}, we apply our formalism to the AB ring system, detailing the U(1)-QHFPE and computing the equilibrium distribution, linear response spectrum, and AB persistent current. Finally, Sec. \ref{sec:Conclution} provides concluding remarks.

\section{Theory}
\label{sec:theory}
\subsection{Model Hamiltonian}
Consider a 3D system with a gauge field that interacts with heat baths. The total Hamiltonian is expressed as
\begin{eqnarray}
\hat{H}_{\rm tot}^{\rm RISB} ( t ) &&=   \hat{H}_S(t) +  \hat{H}_{I+B}^{\rm RISB},
\label{eq:BrownianH}
\end{eqnarray}
where
\begin{eqnarray}
\hat H_{S} ( t ) = \frac{1}{2m_S} \left[ \hat{\bm{p}} - q \bm{A} ( \hat{\bm{x}} ; t ) \right]^2  + U_\phi ( \hat{\bm{x}} ; t )
\label{eq:system}
\end{eqnarray}
is the Hamiltonian of the system (subsystem) with mass $m_S$ described by the momentum $\hat {\bm p}$ and the coordinate $\hat {\bm x}$. The system with charge $q$ interacts with the vector potential and scalar potential expressed as ${\bm A}(\hat{\bm x}; t)$ and $\phi(\hat{\bm x}; t)$; we treat the scalar potential as part of the potential $U_{\phi}(\hat{\bm x}; t)$. In the three-dimensional (3D) RISB model, the particle system is independently coupled to three heat baths in the $\alpha=x$, $y$, and $z$ directions through the functions of $\hat V_{\alpha}$. The bath system is then expressed as\cite{IT19JCP}
\begin{eqnarray}
\hat{H}_{I + B}^{\rm RISB} &&\nonumber \\
= \sum_{\alpha}^{x,y,z}&& \sum_k \left\{
\frac{(\hat{p}_{k}^{\alpha})^2}{2 m_k^{\alpha}}  + \frac{m_k^{\alpha}  (\omega_k^{\alpha})^2}{2} \left(\hat{q}_k^{\alpha}  -
\frac{c_{k}^{\alpha} \hat V_{\alpha}}{m_{k}^{\alpha} (\omega_{k}^{\alpha})^{2}} \right)^2\right\}, \nonumber \\
\label{eq:Balpha}
\end{eqnarray}
where $m_k^{\alpha}$, $\hat{p}_k^{\alpha}$, $\hat q_k^{\alpha}$ and $\omega_k^{\alpha}$ are the mass, momentum, position, and frequency variables of the $k$th bath oscillator mode, and $c_k^{\alpha}$ is the S-B coupling constant in the $\alpha$ direction. If necessary, we can set anisotropic environments, for example in the $xy$ direction by correlating the bath modes in the $\alpha=x$ and $y$ direction.\cite{TT20JPSJ} We have introduced the counter terms $\sum_k (c_k^{\alpha})^2\hat{V}_{\alpha}^2  /2m_k^{\alpha} ( \omega_k^\alpha )^2$ to maintain the translational symmetry of the Hamiltonian in the $\alpha$ directions.\cite{TW91PRA,TW92JCP,T15JCP,T06JPSJ,T20JCP} 
In the 2D angular coordinate representations $(r, \theta)$,  $\hat V_{\alpha}$ are expressed as $\hat{V}_x = r \cos({\theta})$ and $\hat{V}_y = r \sin({\theta})$,\cite{IT18JCP} whereas in the 3D angular coordinate representations $(r, \theta, \phi)$, they are expressed as $\hat{V}_x = r \sin({\theta}) \cos( \phi)$, $\hat{V}_y = r \sin({\theta})\sin(\phi)$, and 
$\hat{V}_z = r \cos({\theta})$,\cite{IT19JCP,Lipeng3Drotar2019}
 respectively. 

The heat bath is characterized by the spectral distribution function (SDF) defined as
$J^{\alpha}(\omega) =\sum_k [{{\hbar}( c_k^{\alpha})^2}/{2m_k^{\alpha} \omega_k^{\alpha}}] \delta(\omega - \omega_k^{\alpha})$,
and the inverse temperature, $\beta \equiv 1/k_{\mathrm{B}}T$, where $k_\mathrm{B}$ is the Boltzmann constant. 
It should be noted that $J^{\alpha}(\omega)$ for different $\alpha$ need not be the same, when the surrounding environment is anisotropic.

For the collective coordinate of the bath mode in the $\alpha$ direction expressed as $\hat X_{\alpha} \equiv \sum_k c_k^{\alpha} \hat x_k^{\alpha}$, the subsystem is considered to be driven by  the external force $\hat X_{\alpha} (t)$ through the interaction $-V_{\alpha} (\hat{\bm{x}}) \hat X_{\alpha} $,
where $\hat{X}_{\alpha} (t)$  
is the Heisenberg representation of $\hat{X}_{\alpha} $ for the bath Hamiltonian $\hat H_B^{\alpha} = 
\sum_j \left( ( \hat{p}_j^{\alpha} )^2 / {2 m_j^{\alpha} } + m_j^{\alpha}  (\omega_j^{\alpha} )^2 (\hat x_j^{\alpha} )^2 / 2 \right)$.\cite{TK89JPSJ1,T06JPSJ}
Since the bath is harmonic, it has a Gaussian nature whose thermal equilibrium state is described by the Boltzmann distribution $\exp[-\beta H_B^{\alpha} ]$.\cite{T06JPSJ} Thus, the character of 
$\hat X_{\alpha} (t)$ is specified by its two-time correlation functions, such as
the canonical and symmetrized correlation functions defined as $\Psi_{\alpha} (t) = \beta \langle \hat{X}_{\alpha}  ; \hat{X}_{\alpha} (t) \rangle_\mathrm{B}$ and $C_{\alpha} (t) = 
\langle
\hat{X}_{\alpha} (t)\hat{X}_{\alpha} (0)+\hat{X}_{\alpha} (0)\hat{X}_{\alpha} (t)\rangle_\mathrm{B}/2$,
where $\langle \cdots \rangle_\mathrm{B}$ represents the thermal average of the bath degrees of freedom. 
The function $C_{\alpha} (t)$ is analogous to the classical correlation function of $X_{\alpha} (t)$.  
The effect of a heat bath consists of fluctuation $C_{\alpha} (t)$ and dissipation $\Psi_{\alpha} (t) $, which satisfies the fluctuation-dissipation relation;\cite{TK89JPSJ1,T06JPSJ}  
$C_{\alpha} [\omega] = \hbar \omega \coth(\beta \hbar \omega/2)  \Psi_{\alpha} [\omega]/2$.
In terms of $J_{\alpha} (\omega)$, they are expressed as 
\begin{eqnarray}
\Psi_{\alpha} (t) = \frac{2}{\hbar}\int_0^\infty  {d\omega \frac{{J_{\alpha} (\omega )}}
{\omega }\cos \left( {\omega t} \right)} ,
\label{eq:L_1barDef}
\end{eqnarray}
and
\begin{eqnarray}
C_{\alpha} (t) = \int_0^\infty  {d\omega J_{\alpha} (\omega ) \coth \left( {\frac{{\beta \hbar \omega }}{2}} \right) \cos \left( {\omega \,t} \right)}. 
\label{eq:L_2Def}
\end{eqnarray}
This is also equivalent to the kernel function, $L^\alpha(t) = iL^\alpha_1 (t) +L^\alpha_2 (t)$, appearing in the Feynman-Vernon influence functional [{\it i.e.} $- \partial \Psi_\alpha ( t ) / \partial t = 2 L_1^\alpha ( t )/ \hbar $ and $C_\alpha(t) = L^\alpha_2 (t)$].\cite{feynman1963dynamical,T14JCP}

We assume that $J^\alpha ( \omega )$ has an Ohmic form with a Lorentzian cut-off (Drude SDF) expressed as\cite{T06JPSJ,T20JCP}
\begin{eqnarray}
 J^{\alpha}(\omega) = \frac{\hbar \eta_{\alpha}}{\pi}\frac{\gamma_{\alpha}^2\omega}{\gamma_{\alpha}^2+\omega^2},
\label{JDrude}
\end{eqnarray}
where the constant $\gamma_{\alpha}$ represents the width of the spectral distribution of the collective 
bath modes and is the reciprocal of the correlation time of the bath-induced noise in the $\alpha$ direction. 
The parameter $\eta_{\alpha}$ is the S-Bcoupling strength, which represents the magnitude of damping. This spectral distribution approaches the Ohmic distribution, $J^{\alpha}(\omega)\approx \hbar \eta_{\alpha}\omega /\pi$, for large $\gamma_{\alpha}$. 

For the Drude SDF, using the Pad{\'e} frequency decomposition, the kernel function $L^\alpha ( t )$ is approximated as 
\begin{eqnarray}
\label{eq:KernelPade}
L^\alpha ( t ) &\simeq& i \frac{\hbar \eta_\alpha \gamma_\alpha^2}{2} e^{- \gamma_\alpha | t |} \nonumber \\
&+& \frac{\gamma_\alpha}{\beta} \left( 1 
+ \sum_{j = 1}^{K_\alpha} \frac{2 \bar{\eta}_j^\alpha \gamma_\alpha^2}{\gamma_\alpha^2 - (\nu_j^{\alpha})^2} \right)
e^{- \gamma_\alpha | t |} \nonumber \\
&+& \sum_{j = 1}^{K_\alpha} \left[ - \frac{2 \eta_\alpha \gamma_\alpha^2}{\beta}
\frac{\bar{\eta}_j^\alpha \nu_j^{\alpha}}{\gamma_\alpha^2 - (\nu_j^{\alpha})^2} e^{- \nu_j | t |} \right] ,
\end{eqnarray}
where $\nu_j^{\alpha}$ and $\bar{\eta}_j^{\alpha}$ are the $j$th Pad{\'e} frequency and coefficient and 
$K_{\alpha}$ is the number of the Pad{\'e} elements in the $\alpha$ direction.\cite{hu2010communication}

\subsection{The U(1)-HEOM}
 
For the Hamiltonian Eqs. \eqref{eq:BrownianH}-\eqref{eq:Balpha}, the HEOM are expressed as\cite{T14JCP,T06JPSJ,T20JCP,Lipeng3Drotar2019}
\begin{eqnarray}
\label{eq:HEOM}
\frac{\partial \hat{\rho} _{\{ \bm{n}_\alpha \}} ( t ) }{\partial t} 
&& = - \left[ \frac{i}{\hbar} \hat H_S^\times ( t )  + \sum_\alpha^{x , y , z} 
\sum_{j = 0}^{K_\alpha} n_j^\alpha \nu_j^{\alpha}  \right] \! \hat{\rho}_{ \{ \bm{n}_\alpha \} } ( t ) 
\nonumber \\
&&
+ \sum_\alpha^{x , y , z} \sum_{j = 0}^{K_\alpha} \hat{\Phi}^\alpha 
\hat{\rho} _{\{ \bm{n}_\alpha + \bm{e}_\alpha^j \}} ( t ) \nonumber \\
&&
+ \sum_\alpha^{x , y , z} \sum_{j  = 0}^{K_\alpha} n_j^\alpha \hat{\Theta}_j^\alpha
\hat{\rho}_{\{ \bm{n}_\alpha - \bm{e}_\alpha^j \}} ( t ) ,
\end{eqnarray}
where ${\{ \bm{n}_\alpha \}} \equiv ( \bm{n}_x , \bm{n}_y , \bm{n}_z )$  is a set of integers 
$\bm{n}_\alpha = ( n_\alpha^0 , n_\alpha^1 , n_\alpha^2 , \cdots,  n_\alpha^{K_\alpha} )$
to describe the hierarchy elements and ${\{ \bm{n}_\alpha \pm \bm{e}_\alpha^j \}}$, where the index $j \in \{ 0 , 1 , \cdots , K_\alpha \}$,
 represents, for example, ${\{ \bm{n}_\alpha \pm \bm{e}_y^j \}} = ( \{ \bm{n}_x \} , n_y^0 , \cdots, n_y^j \pm 1 , \cdots, \{ \bm{n}_z \} )$ for $\alpha = y$. In Eq.~\eqref{eq:HEOM}, we set $\nu_\alpha^0 = \gamma_\alpha$.  We employed the hyper operators defined as $\hat{\mathcal{O}}^\times \hat f \equiv \hat{\mathcal{O}} \hat f - \hat f \hat{\mathcal{O}}$ and $\hat{\mathcal{O}}^\circ \hat f \equiv \hat{\mathcal{O}} \hat f + \hat f \hat{\mathcal{O}}$ for any operator $\hat{\mathcal{O}}$ and $\hat{f}$. The relaxation operators are then expressed as
$\hat{\Phi}^\alpha \equiv -i \hat{V}_\alpha^\times / \hbar$,
\begin{eqnarray}
\label{eq:Theta_S}
 \hat{\Theta}_0^{\alpha} ( t ) \equiv~~~~~~~~~~~~~~~~~~~~~~~~~~~~~~~~~~~~~~~~~~~~~~~~~~~~~~~~~~~~~  \nonumber \\
 - \frac{i}{\hbar} 
\left[ \frac{\gamma_\alpha \eta_\alpha}{\beta}
\left( 1 + \sum_{j = 1}^{K_\alpha} \frac{2 \bar{\eta}_j^\alpha \gamma_\alpha^2}{\gamma_\alpha^2 - (\nu_j^{\alpha})^2} \right)
\hat{V}_\alpha^\times + \frac{i \hbar \eta_\alpha \gamma_\alpha}{2} \hat{\dot{V}}_\alpha^\circ ( t ) \right] ,\nonumber \\
\end{eqnarray}
and
\begin{eqnarray}
\label{eq:Theta_0}
\hat{\Theta}_j^\alpha \equiv \frac{i}{\hbar} \frac{\gamma_\alpha^2 \eta_\alpha \bar{\eta}_j^{\alpha} }{\beta}
\frac{2 \nu_j^{\alpha}}{\gamma_\alpha^2 - (\nu_j^{\alpha})^2} \hat{V}_\alpha^\times .
\end{eqnarray}
We set $\hat \rho_{\{ \bm{n}_\alpha - \bm{e}_\alpha^j \}} ( t ) = 0$ when $\bm{n}_\alpha = ( \cdots , n_\alpha^j = 0 , \cdots )$.  
Note that $\hat{\dot{V}}_\alpha ( t ) \equiv i [ \hat{H}_S ( t ) , \hat{V}_\alpha ] / \hbar$ in Eq.~\eqref{eq:Theta_S} appears because we made the integration by parts to eliminate the counter term of the total Hamiltonian utilizing the Heisenberg equation of motion, $i.e.-i\hbar\dot{\mathcal{O}_{\rm H}}(t) = [H_S (t), \mathcal{O}_{\rm H}(t)]$, as in the case of the quantum Fokker--Planck equation.\cite{T06JPSJ,T20JCP,T15JCP,TW91PRA,TW92JCP}  

The Markov approximation is achieved by taking two limits: (i) the short noise correlation limit ($\gamma_\alpha \gg \omega_0$), where $\omega_0$ is the characteristic frequency of the system, and (ii) the high temperature limit ($\beta \hbar \gamma_\alpha/2 \ll 1$).\cite{TK89JPSJ1,IT19JCTC,T06JPSJ,T20JCP,T15JCP}   We then obtain the quantum Master equation (QME) as\cite{T06JPSJ,T20JCP,T15JCP}
\begin{align}
\frac{\partial \hat{\rho}_0 ( t )}{\partial t}  &= - \frac{i}{\hbar} \hat{H}_S^\times ( t ) \hat{\rho}_0 ( t )\nonumber \\
&
- \sum_\alpha^{x , y , z} \frac{\eta_\alpha}{\hbar^2} \hat{V}_\alpha^\times 
\left[ \frac{1}{\beta} \hat{V}_\alpha^\times + \frac{i}{2} \hat{\dot{V}}_\alpha^\circ ( t ) \right] 
\hat{\rho}_0 ( t ) .
\label{eq:MasterEq}
\end{align}
Note that to satisfy conditions (i) and (ii), the bath temperature must be extremely high ($\beta\hbar\ll1/\gamma_{\alpha}\approx0$). Therefore, {\it the Markov limit is unphysical at low bath temperatures,} even when the rotating wave approximation (RWA) and factorization assumption (FA) are applied.\cite{Weiss2012,KT24JCP3,KT24JCP4}
Therefore, except for the free rotor,\cite{ST02JPSJ,ST03JCP,IT18JCP} the above equation {\it cannot describe many important quantum effects,} where the system and bath are entangled and non-RWA terms play an important role. 
In the classical limit $\hbar \rightarrow 0$, this issue does not occur.

A particle moving in an electromagnetic field is expected to satisfy the unitary group symmetry [U(1)] with gauge invariance. This requirement is essential even when the particles are in a heat bath, and the reduced equations of motion that are derived from the gauge-invariant Hamiltonian Eqs. \eqref{eq:BrownianH}-\eqref{eq:Balpha}.
Thus, the HEOM, Eqs.  \eqref{eq:HEOM}-\eqref{eq:Theta_0}, must satisfy gauge invariance for
${\bm A} ( {\bm x} ; t )$ and $-\phi( {\bm x} ; t)$.

To demonstrate this point, we consider the gauge transformation ${\bm A'}({\bm x}; t) = {\bm A}({\bm x}; t)+\nabla\chi({\bm x},t)$ and $\phi' ({\bm x}; t)= \phi({\bm x}; t) -\partial_t \chi({\bm x},t)$, where $\chi({\bm x},t)$ is a differentiable arbitrary gauge field with the state $| \Psi ( t ) \rangle$ is transformed as $| \Psi' ( t ) \rangle = \hat{U} ( \hat{\bm{x}} , t ) | \Psi ( t ) \rangle$, where $\hat{U} ( \hat{\bm{x}} , t ) \equiv e^{i q \chi( \hat{\bm{x}} , t ) / \hbar}$ is the Unitary operator, which satisfies 
\begin{eqnarray}
\partial_{\alpha'} U ( \hat{\bm{x}} , t ) = i q \left( \partial_{\alpha'} \chi( \hat{\bm{x}} , t ) \right) U ( \hat{\bm{x}} , t ) .
\label{eqn:U1}
\end{eqnarray}
Here the index $\alpha'=0$ is time component, $\alpha'=1-3$ are those for the space components, and
\begin{align}
\frac{\partial}{\partial t} U ( \hat{\bm{x}} , t ) | \Psi ( t ) \rangle &= U ( \hat{\bm{x}} , t ) 
\frac{\partial}{\partial t} | \Psi ( t ) \rangle \nonumber \\
&+ i q \frac{\partial \chi ( \hat{\bm{x}} , t )}{\partial t} U ( \hat{\bm{x}} , t ) | \Psi ( t ) \rangle .
\label{eqn:U2}
\end{align}
The gauge invariance of the U(1)-HEOM for the above transformation is presented in Appendix \ref{Sec:GIHEOM}.

\subsection{The U(1)-QHFPE}
The HEOM in the Wigner space is ideal for studying quantum transport systems, because it allows the treatment of continuous systems, utilizing open boundary conditions\cite{ST14NJP} and periodic boundary conditions.\cite{TW92JCP,KT13JPCB,IDT19JCP} When the physical length of the subsystem is longer than its quantum coherence, the rotational system can be handled using the periodic boundary conditions of the Wigner distribution function (WDF). Nevertheless, a particular consideration is necessary when the physical dimensions of the subsystem are much smaller than the coherence, as is the case with the nanoscale system.

We thus start from a periodic system with length $L_{\alpha}$ and introduce the discretized Wigner transformation with periodic boundary conditions (DWT-PBC). For the density operator $\rho_{\{{\bm n}_{\alpha}\}}\left({\bm x}, {\bm x}' ; t\right) ,$ this is defined as
\begin{eqnarray}
W_{\{ \bm{n}_\alpha \}} ( \bm{p}_n , \bm{r} ) 
&&= \prod_\alpha^{x , y , z} \left[ \frac{1}{2 \pi \hbar} \int^{ L_\alpha}_{- L_\alpha} 
d \xi_\alpha e^{- \frac{i p^\alpha_n \xi_\alpha}{\hbar}} \right] \nonumber \\
&&\times \rho_{\{ \bm{n}_\alpha \}} \left( \bm{r} + \frac{\bm{\xi}} 2 , \bm{r} - \frac{\bm{\xi}} 2 \right) ,
\label{eqn:DisWigTran}
\end{eqnarray}
where $\xi_\alpha\in[ - L_\alpha ,  L_\alpha ]$ and $L_\alpha$ is the system size in the $\alpha$ direction.
The discretized momentum is represented as $\bm{p}_n = ( p^x_n, p^y_n, p^z_n )$, where each grid point is defined as $p^\alpha_n \equiv n^\alpha\Delta p^\alpha$, with $\Delta p^\alpha \equiv \pi \hbar/ L_\alpha$ for an integer $n^\alpha$.  Note that the integral region $L_{\alpha}$ with respect to $\xi_\alpha$ is twice the period, due to the factor of $1 / 2$ applied to $\bm{\xi}$.
Accordingly, the inverse Wigner transformation is defined as 
\begin{align}
\rho_{\{{\bm n}_{\alpha}\}}\left({\bm x}, {\bm x'} \right) = 
\prod_\alpha^{x,y,z}\left[ \Delta p^\alpha \sum_{\alpha=-\infty}^{\infty} e^{\frac{ip^\alpha_n\xi_\alpha}{\hbar}} \right]W_{\{{\bm n}_{\alpha}\}}( \bm{p}_n , \bm{r} ) 
\end{align}
with ${\bm x} \equiv {\bm r}+{\boldsymbol\xi}/2$ and $ {\bm x}' \equiv{\bm r}-{\boldsymbol\xi} /2$.
The normalization condition of the DWF is defined as
\begin{align}
\prod_\alpha^{x , y , z} \left[ \int_{- \frac{L_\alpha}{2}}^{ \frac{L_\alpha}{2}} d r_\alpha
\sum_{n^\alpha = - \infty}^\infty \Delta p^\alpha \right] W_{\{ \bm{0} \}} ( \bm{p}_n , \bm{r} ) = 1 ,
\end{align}
where $W_{\{ \bm{0} \}}( \bm{p}_n , \bm{r} ) $ is the auxiliary WDF with $\bm{n}_x = \bm{0} , \bm{n}_y = \bm{0} ,$ and $\bm{n}_z = \bm{0}$.

When $L_\alpha$ is larger than the coherent length of the subsystem, we can take the continuous space limit that satisfies the Riemann integral as
\begin{eqnarray}
\label{Wigner}
\lim_{L_\alpha \rightarrow \infty} \sum_{n^\alpha = - \infty}^{ \infty} \Delta p^\alpha
= \int_{- \infty}^{+ \infty} d p^\alpha .
\end{eqnarray}
As a result, the integral region in the Wigner transformation is modified from the discrete momentum space $ \bm p_n$ to the continuous momentum space $\bm p$. Consequently, the Wigner transformation adheres to its standard definition expressed as\cite{T06JPSJ,T20JCP,T15JCP,TW91PRA,TW92JCP,CALDEIRA1983587,WaxmanFP1985}
\begin{eqnarray}
W_{\{ \bm{n}_\alpha \}}  ( \bm{p} , \bm{r} ) && \equiv \prod_\alpha^{x , y , z} 
\left( \frac{1}{2 \pi \hbar} \int_{- \infty}^{\infty} d {\xi_\alpha} e^{-\frac{i p_\alpha \xi_\alpha}{\hbar}}
\right)  \nonumber \\
& &
~~~~~~\times \rho_{\{ \bm{n}_\alpha \}} \left( \bm{r} + \frac{\bm{\xi}}{2} , \, \bm{r} - \frac{\bm{\xi}}{2} \right).
\label{eqn:WigTran}
\end{eqnarray}
In this case, Eq. \eqref{eq:HEOM} are rewritten as
\begin{align}
\frac{\partial W_{\{ \bm{n}_\alpha \}} ( \bm{p} , \bm{r})}{\partial t}  &= 
- \left[ \hat{\mathcal{L}}_{qm} (t)
+ \sum_\alpha^{x , y , z} \sum_{j = 0}^{K_\alpha} n_\alpha^j \nu_j^\alpha \right] \nonumber \\
& ~~~~ \times W_{\{ \bm{n}_\alpha \}} ( \bm{p} , \bm{r} ) \nonumber \\
&
~~~~ + \sum_\alpha^{x , y , z} \sum_{j = 0}^{K_\alpha}
\tilde{\Phi}^\alpha W_{\{ \bm{n}_\alpha + \bm{e}_\alpha^j \}} ( \bm{p} , \bm{r} )  \nonumber \\
&
~~~~ + \sum_\alpha^{x , y , z}  \sum_{j = 0}^{K_\alpha}
n_\alpha^j \tilde{\Theta}_j^\alpha (t)W_{\{ \bm{n}_\alpha - \bm{e}_\alpha^j \}} ( \bm{p} , \bm{r} ).
\label{heom_wig}
\end{align}
Here, the quantum Liouvillian with gauge field for $\hat H_S (t)=H_S (\hat {\bm p}, \hat {\bm x} ; t)$ in 
Eq. \eqref{eq:system} is expressed in integral form by extending Frensley's formalism as\cite{Frensley1990}
\begin{align}
- \mathcal{\hat L}_{qm} ( t ) W_{\{ \bm{n}_\alpha \}}  ( \bm{p} , \bm{r} ) &
=-\frac{i}{\hbar} \prod_\alpha^{x , y , z} \left( \frac{1}{( 2 \pi \hbar )^2} 
\int_{-\infty}^{\infty} d p'_\alpha d r'_\alpha \right) \nonumber \\
&
\times
T_{S} ( \bm{p} , \bm{r} , \bm{p}' , \bm{r}' ; t )
W_{\{ \bm{n}_\alpha \}}  ( \bm{p}' , \bm{r}' ) ,
\label{eq:quantumLioiv}
\end{align}
where the integral kernel is defined as
\begin{align}
& T_S ( \bm{p} , \bm{r} , \bm{p}' , \bm{r}' ; t ) \nonumber \\
&~~\equiv \prod_\alpha^{x , y , z} \left( \int^\infty_{- \infty} d \xi_\alpha d \zeta_\alpha \right)
e^{- \frac{i ( \bm{p} - \bm{p}' ) \cdot \bm{\xi}}{\hbar}}
e^{\frac{i ( \bm{r} + \bm{r}' ) \cdot \bm{\zeta}}{\hbar}} \nonumber \\
&
~~\times
\left\{ H_{S}^\rightarrow \left( \bm{p}' + \frac{\bm{\zeta}}{2} , \bm{r} + \frac{\bm{\xi}}{2} ; t \right)
- H_{S}^{\leftarrow} \left( \bm{p}' - \frac{\bm{\zeta}}{2} , \bm{r} - \frac{\bm{\xi}}{2} ; t \right) \right\} .
\end{align}
Here, $H_{S}^\rightarrow$ and $H_{S}^\leftarrow$ represent the system Hamiltonian, with the momentum operators positioned to the right and left of the density operator, respectively. In arranging them, it is essential to account for the commutation relation between the position and momentum operators.

The auxiliary operators are expressed as $\tilde{\Phi}^{\alpha}  =  {\partial }/{{\partial p_{\alpha} }}$,  
\begin{eqnarray}
\tilde{\Theta}_0^\alpha ( t ) && 
\equiv \frac{\eta_\alpha \gamma_\alpha}{m_{S}}
\Bigg[ \left( p_\alpha - q \hat{A}_\alpha ( t ) \right) \nonumber \\
&&
\quad
+ \frac{m_{S}}{\beta}
\left( 1 + \sum_{j = 1}^{K_\alpha} 
\frac{2 \bar{\eta}_j^\alpha \gamma_\alpha^2}{\gamma_\alpha^2 - ( \nu_j^\alpha )^2} \right)
\frac{\partial}{\partial p_\alpha} \Bigg] , 
\label{eqn:WigTheta_S}
\end{eqnarray}
and
\begin{equation}
\tilde{\Theta}_j^\alpha = - \frac{\eta_\alpha \gamma_\alpha^2}{\beta}
\frac{2 \bar{\eta}_j^\alpha \nu_j^\alpha}{\gamma_\alpha^2 - ( \nu_j^\alpha )^2} \frac{\partial}{\partial p_\alpha} 
\quad ( j = 1 , \cdots , K_\alpha ) .
\end{equation}
Here, we introduce the operator defined as
\begin{align}
\label{eq:VectorPotentialOperator}
& \hat{A}_\alpha ( t ) W ( \bm{p} , \bm{r} )
= \frac{1}{2 ( 2 \pi \hbar )^3} \int d \bm{p}' \int d \bm{\xi} 
e^{\frac{i \bm{\xi} \cdot ( \bm{p}' - \bm{p} )}{\hbar}}
\nonumber \\
& \quad \times \left\{ A_\alpha \left( \bm{r} + \frac{\bm{\xi}}{2} , t \right)
+ A_\alpha \left( \bm{r} - \frac{\bm{\xi}}{2} , t \right) \right\} W \left( \bm{p}' , \bm{r} \right) .
\end{align}

The U(1)-QHFPE satisfies gauge invariance as in the case of U(1)-HEOM presented in Eqs. \eqref{eq:HEOM}-\eqref{eq:Theta_0}. For the damping operator in Eq. (\ref{eqn:WigTheta_S}), this is obvious because it involves the kinetic momentum ${{\bm p}-q\hat {\bm A}(t)}$ instead of the canonical momentum ${\bm p}$. 

In the Markovian limit, we have the U(1) quantum Fokker--Planck equation (U(1)-QFPE) expressed as
\begin{eqnarray}
&&\frac{\partial W ( \bm{p} , \bm{r})}{{\partial t}}=  - \mathcal{\hat{L}}_{qm}(t) W ( \bm{p} , \bm{r} ) \nonumber \\
&&~~~~ +\sum_\alpha^{x,y,z}  \frac{ \eta_\alpha}{m_{S}} \frac{\partial}{\partial p_\alpha}
 \left\{ \left[ p_\alpha - q \hat{A}_\alpha ( t ) \right] 
+ \frac{m_{S}}{\beta} \frac{\partial}{\partial p_\alpha} \right\}
 W ( \bm{p} , \bm{r} ).\nonumber \\
\label{eqn:U(1)FokkerPlanck}
\end{eqnarray}
This expression is valid in high-temperature semi-classical conditions. The above equation is equivalent to the Caldeira--Leggett quantum Fokker--Planck equation (CL-QFPE) when $q A_\alpha = 0$ and $q U_\phi = 0$.\cite{T06JPSJ,T15JCP,TW91PRA,TW92JCP,CALDEIRA1983587,WaxmanFP1985} 

Because the high-temperature limit $\beta \hbar \ll 1/\gamma \approx 0$ is equivalent to the semiclassical limit, $\hbar \rightarrow 0$, the quantum Liouvillian is approximated as
\begin{align}
- \mathcal{\hat{L}}_{\rm cl} (t)
= \sum_\alpha^{x,y,z} \left[ -\frac{ [p_\alpha - q \hat{A}_\alpha ( t ) ] }{m_S}
\frac{\partial}{\partial r_\alpha} + \frac{\partial U_\phi ( \bm{r} , t )}{\partial r_\alpha}
\frac{\partial }{\partial p_\alpha} \right]. & \nonumber \\
\end{align}

\section{Numerical Demonstration: AB ring system}
\label{sec:ABring} 

\begin{figure}
    \centering
    \includegraphics[width=1\linewidth]{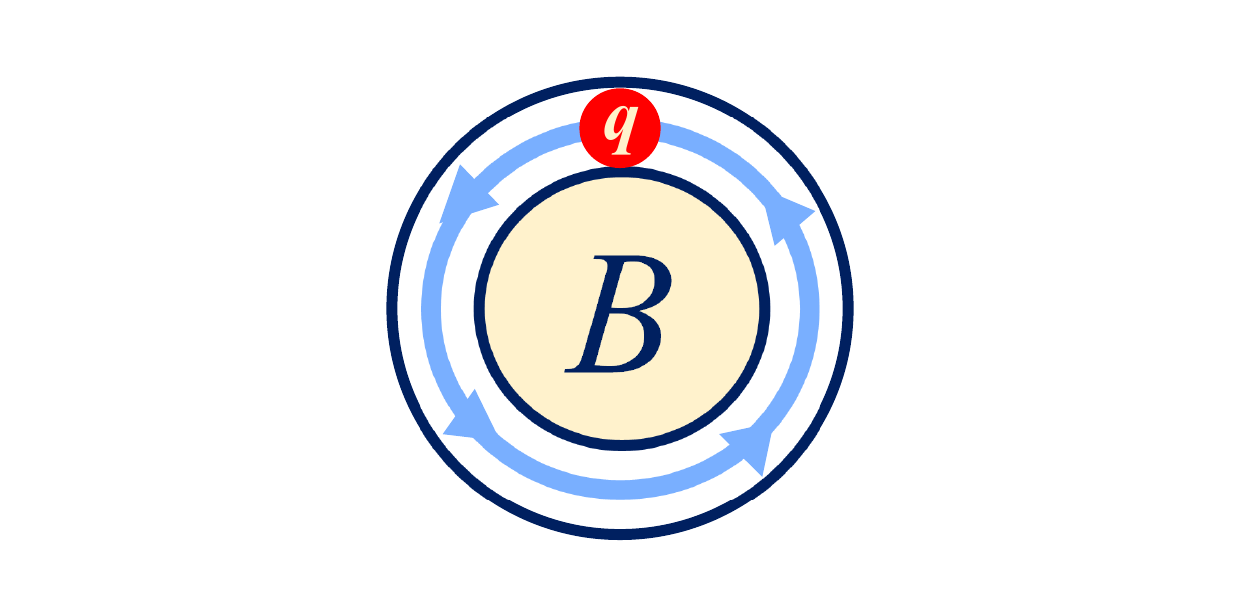}
    \caption{Schematic of an AB ring system. We consider the single flow case, in which a single charge rotates in one direction along the ring. The magnetic field is confined only inside the ring and is set to zero outside, as shown in Table \ref{tab:MagVec}.}
\label{fig:ABring}
\end{figure}

The AB effect is a phenomenon that arises from the presence of an electromagnetic gauge field with U(1) symmetry in a quantum system.\cite{1959Aaronov-Bohm,Tonomura1986PhysRevLett.56.792,2008NoriTonomura}  To observe this effect, we consider a simple model known as the AB ring.\cite{ByersYang1961,1986HalperinRevModPhys.58.533,1991ii,Blasi_ACring2023,1986HalperinRevModPhys.58.533,1991ii,1992FukuyamaAndo,BookImry,1986ABPhysRevLett.56.386,1987Umbach10.1063/1.97887,levy1990magnetization,bluhm2009persistent,PC2010,Fomin2018} This system is described by the Hamiltonian Eq. \eqref{eq:system} with the potential $U(\hat{\bm x}; t)=0$.
As problems of the AB ring, here we consider a single charge rotating around the ring, and there is a vertical magnetic field perfectly shielded at the center of the ring.\cite{1992FukuyamaAndo,BookImry,1986ABPhysRevLett.56.386,1987Umbach10.1063/1.97887,levy1990magnetization,bluhm2009persistent,PC2010,Fomin2018} (See Fig. \ref{fig:ABring}). The magnetic and vector fields are set as shown in Table \ref{tab:MagVec}.

\renewcommand{\arraystretch}{2} 
\begin{table}
    \centering
    \begin{tabular}{||c|c|c||}
         \hline
         & $\vec B$ & $\vec A$ \\
         \hline
        Inside ($r < r_0$) & $ \quad B_0 \hat z \quad $ & $\frac{B_0}{2} r \hat \theta$ \\
        \hline
        \ Outside ($r \ge r_0$) \ & $0$ & $ \quad \frac{B_0 r_0^2}{2 r}\hat \theta = \frac{\Phi}{2\pi r}\hat \theta \quad $ \\
        \hline
    \end{tabular}
    \caption{Magnetic fields and vector potentials set up in the A--B ring of radius $r_0$. Here, the magnetic field inside the ring is constant at magnitude $B_0$ and the magnetic flux is $\Phi = \pi r_0^2 B_0$, while the field outside the ring is set to zero. Outside of it, the magnetic field is completely shielded, but the vector field is not negligible.}
    \label{tab:MagVec}
\end{table}

The total Hamiltonian of the AB ring is then expressed as
\begin{align}
\label{eqn:AB}
&\hat H_{tot}^{\rm RISB} = \frac{ (\hat p_\theta - q r_0 A )^2}{2 I_S} \nonumber \\
&~~~~~~~+ \sum_{\alpha}^{x,y}\sum_{k}\left[\frac{(\hat p_k^\alpha)^2}{2m_k^\alpha}+\frac{m_j^\alpha\omega_k^\alpha}{2}\left(\hat q_k^\alpha - \frac{c_k^\alpha \hat V_\alpha}{m_k^\alpha (\omega_k^\alpha)^2}\right)\right],
\end{align}
where $\hat p_\theta$ is the angular momentum, $I_S \equiv m_S  r_0^2$ is the moment of inertia, and $\hat V_x = r_0 \cos\hat\theta$ and $\hat V_y = r_0 \sin\hat\theta$.

Note that the Wigner transformation for the 2D polar coordinate system is not uniquely defined,\cite{nieto1998wigner,IT21JCEL} but here we adopt the DWT-PBC, as specified in Eq. \eqref{eqn:DisWigTran}, to conduct numerical simulations.

For the Drude environment Eq. \eqref{JDrude}, we consider the isotropic bath $J^{x}(\omega)= J^{y}(\omega)= J(\omega)$.
Thus, the U(1)-QHFPE for the AB ring are expressed as
\begin{eqnarray}
\label{eq:QHFPE-C}
\frac{\partial W_{\{ \bm{n}_\alpha \}} (p_n, \theta  ) }{\partial t}&=&
 - \left[ \hat{\mathcal{L}}_{qm}^{\rm AB}
+ \sum_\alpha^{x , y} \sum_{k = 0}^K n_k^\alpha \nu_k^\alpha \right]  \nonumber \\
&& ~~~~ \times W_{\{ \bm{n}_\alpha \}} ( p_n, \theta  ) 
\nonumber \\
&&
+ \sum_\alpha^{x , y} \sum_{k = 0}^K \tilde{\Phi}_\alpha^{\rm AB} 
W_{\{ \bm{n}_\alpha + \bm{e}_\alpha^k \}} ( p_n, \theta  ) \nonumber \\
&&
+ \sum_\alpha^{x , y} \sum_{k = 0}^K n_k^\alpha \tilde{\Theta}_{\alpha , k}^{\rm AB}
W_{\{ \bm{n}_\alpha - \bm{e}_\alpha^k \}} ( p_n, \theta  ) ,\nonumber \\
\end{eqnarray}
where $W_{\{ \bm{n}_\alpha \}} ( p_n, \theta  )$ is an auxiliary WDF for the discretized momentum $p_n$ and
the quantum Liouvillian for the AB ring system is defined as
\begin{eqnarray}
-\mathcal{L}_{qm}^{\rm AB}
W (p_n, \theta ) &=& -\frac{(p_n - q r_0 A )}{I_{S}}\frac{\partial W (p_n, \theta )}{\partial \theta} \nonumber \\
&&- \frac{1}{\hbar} \sum_{k=1}^\infty 
\left[ u_k^{\rm ( c )} \sin ( k \theta ) - u_k^{\rm ( s )} \cos ( k \theta ) \right] \nonumber \\
&&\times \Big[ W(p_{n+k},\theta) - W(p_{n-k},\theta) \Big],
\end{eqnarray}
where $u_k^{(c)}$ and $u_k^{(s)}$ represent the Fourier coefficients of the $k$-th cosine and sine functions, respectively.

We define $\tilde{\Phi}_\alpha^{\rm AB}  = r_0 f_\alpha ( \theta ) \delta / \delta p_n$, where $\delta / \delta p_n$ is defined as $\delta / \delta p_n f ( p_n ) = ( f ( p_{n + 1} ) - f ( p_{n - 1} ) ) / \hbar$ for an arbitrary function $f ( p_n )$ of $p_n$. The other operators appearing in Eq.~\eqref{eq:QHFPE-C} are defined as
\begin{align}
\label{eq:defOperatorC1}
&\quad \tilde{\Theta}_{\alpha , 0}^{\rm AB} W ( p_n, \theta ) =~~~~~~~~~~~~~~~~~~~~~~~~~~~~~~~~~~~~~~~~~~~~~~~~~~~~~~~~~~~~~~~\nonumber \\
&~~~~~~~~~~~~
\frac{\eta r_0 \gamma}{\beta}\left( 1 + \sum_{j = 1}^K \frac{2 \bar{\eta}_j \gamma^2}{\gamma^2 - \nu_j^2} \right)  f_\alpha (\theta) \frac{\delta W ( p_n, \theta )}{\delta p_n} \nonumber  \\
&~~~~~~~~~~~~- \frac{\eta \gamma^2 r_0}{2} g_\alpha (\theta) 
\left[ W ( p_{n + 1},\theta, ) + W ( p_{n - 1},\theta, ) \right] ,
\end{align}
and
\begin{eqnarray}
\label{eq:defOperatorC3}
\tilde{\Theta}_{\alpha , j}^{\rm AB}
= - \frac{\eta r_0 \gamma^2}{\beta} \frac{2 \bar{\eta}_j \nu_j}{\gamma^2 - \nu_j^2} \ f_\alpha (\theta)
\frac{\delta}{\delta p_n} \quad ( j = 1 , \cdots , K ) ,\nonumber \\
\end{eqnarray}
where we set $( f_x (\theta), g_x ( \theta ) ) = ( - \sin \theta, \cos \theta)$ and $(f_y (\theta), g_y(\theta)) =(\cos \theta, \sin \theta)$.
Unlike the QHFPE in Eq.~\eqref{heom_wig}, we incorporate the counter term into the system Hamiltonian in the QHFPE in Eq.~\eqref{eq:QHFPE-C} to conduct the reduction regarding the radial degree of freedom. (see Appendix~\ref{QHFPE_ABring} for its derivation).

In the Markovian limit, the above equations reduce to  (U(1)-QFPE)
\begin{eqnarray}
&& \frac{\partial W ( p_n , \theta )}{\partial t} =
- \mathcal{L}_{qm}^{\rm AB} W ( p_n , \theta ) \nonumber \\
&&~~~~~~~~~~~+ \frac{\eta r_0^2}{\beta' \hbar^2}
\left[ W ( p_{n + 2} ,\theta ) - 2 W ( p_n , \theta ) + W ( p_{n - 2} , \theta ) \right] \nonumber \\
&&~~~~~~~~~~~ + \frac{\eta}{2 m_{S} \hbar } \Big[(p_{n + 2} - q r_0 A ) W ( p_{n+2} , \theta )
\nonumber \\
&&~~~~~~~~~~~~~~~~~~~~~~~~- ( p_{n - 2} - q r_0 A ) W ( p_{n-2} , \theta ) \Big] .
\label{eqn:FP_RISB}
\end{eqnarray}
Here, we introduce the effective temperature $\beta' = \beta / \left(1 - \beta \hbar^2 / (2 I_S)\right)$, which originates from the coupling between rotational and radial degrees of freedom (see Appendix~\ref{sec:QFPE-ABring}).
The computational source codes underlying the above formulations are presented in a separate manuscript.\cite{KYT25JCP2} 

To illustrate the importance of rotational symmetry of the total Hamiltonians, we discuss the CL model as a reference system.  
The Hamiltonian of the CL model\cite{CALDEIRA1983587} for the AB ring is expressed as
\begin{eqnarray}
\label{eqn:C-L}
\hat H_{\rm tot}^{\rm CL}
=&& \frac{\big(\hat {p}_{\theta} - q r_0 A \big)^2}{2I_{S}} \nonumber \\
&&+ \sum_{k}\left[\frac{\hat {p}_k^2}{2 m_k} + \frac{m_k\omega_k^2}{2} \left( \hat {q}_k - \frac{c_k r_0 \hat \theta}{m_k \omega_j^2} \right)^2 \right] .
\end{eqnarray}
In the CL model, the S-B interaction is linear in $r_0 \hat{\theta} \sum c_k \hat{q}_k$.  Thus, the influence of the bath on the system angle $\theta$ increases linearly with $\theta + 2 \pi$, thereby breaking rotational symmetry.\cite{ST02JPSJ,ST03JCP,IT18JCP} 
Consequently, the DWT-PBC expressed in Eq. \eqref{eqn:DisWigTran} cannot 
be utilized. Instead, we use the standard Wigner function expressed as $W_{\rm CL}(p,\theta)$ with the transformation defined in Eq. \eqref{eqn:WigTran}.  The system is then modeled with periodic boundary conditions, similar to the classical rotor, ensuring that $W_{\rm CL}(p,\theta) = W_{\rm CL}(p, \theta + 2\pi)$. Such treatment is justified when the ring size is large, rather than by electronic coherence.\cite{TW92JCP,KT13JPCB,IDT19JCP}  In fact, as the ring size $L_{\theta}$ increases, the DWT-PBC approaches the normal Wigner transformation, as shown in Eq. \eqref{Wigner}.

We then consider the Drude SDF defined as $J(\omega) = ({\hbar \eta}/{\pi})({\gamma^2\omega}/{\gamma^2+\omega^2})$. 
The reduced equations of motion for CL model with a gauge field are presented in Appendix \ref{CL_ABring}.  The hierarchical elements in the CL model is expressed as $W_{\rm CL}^{(\bm n)}(p,\theta)$.  We refer to these equations as the CL U(1)-QHFPE, which also satisfy gauge invariance. 

In the Markovian limit, the CL U(1)-QHFPE simplifies to the CL U(1)-QFPE expressed as
\begin{eqnarray}
\frac{\partial W_{\rm CL} (p, \theta )}{\partial t} &=&  - \frac{p - q r_0 A}{I_{S}} \frac{\partial}{\partial \theta}
W_{\rm CL} ( p , \theta ) \nonumber \\
&& + \frac{\eta}{m_S} \frac{\partial}{\partial p} \left[ ( p - q r_0 A ) + \frac{I_{S}}{\beta} \frac{\partial}{\partial p} \right] W_{\rm CL}(p, \theta ) . \nonumber \\
\label{eqn:FP_CL}
\end{eqnarray}
Previously, the differences between the CL and RISB models were examined by solving the quantum master equation based on the eigenfunctions of angular momentum.\cite{IT18JCP}  In this study, we verify this by directly solving the U(1)-QHFPE. Then, add the gauge field to the results and investigate the effect. 

 In general, the majority of quantum effects tend to vanish in Markov limits, requiring high-temperature approximations. However, it has been shown that quantum effects due to periodic boundaries are observed when there is no potential, as in the case of a free particle.\cite{IT18JCP,IT19JCP}  Below, we discuss the differences between CL and the RISB model by solving the equilibrium distribution and absorption spectrum under the Markov limit. Then, we solve the U(1)-QHFPE, which can be studied at any temperature, to simulate the persistent current of the AB ring.

Thus, we employed the U(1)-QFPE [Eq.\eqref{eqn:FP_RISB}] to simulate the rotational band in the linear absorption spectrum, whereas the Persistent Current, a feature exclusive to non-Markovian dynamics, was obtained using the U(1)-QHFPE [Eqs.\eqref{eq:QHFPE-C}-\eqref{eq:defOperatorC3}]. The temperature dependence of the rotational band in the non-Markovian regime is addressed in a separate publication that introduces the software developed in this work.\cite{KYT25JCP2}

\subsection{Equilibrium distributions: Markovian case}

First, we investigate the characteristics of the DWT-PBC. To this end, we solve Eqs. \eqref{eqn:FP_RISB} and \eqref{eqn:FP_CL}, which are obtained under the Markov approximation. Note that a comparative analysis between the CL and RISB models for a damped free rotor system, which is described as Eqs. \eqref{eqn:FP_RISB} and \eqref{eqn:FP_CL} with $A=0$ reveal noteworthy insight: The quantum nature of the system emerges from discretized energy states with the frequency $\omega_0 = \hbar / 2 I_{S}$ resulting from the periodic boundary condition.\cite{IT18JCP,IT19JCP}  In this high-temperature limit $\hbar \omega_0 \ll 1/\beta$, where the dynamics is semiclassical by nature, $\hbar$ remains only in the RISB model and thereby satisfying the fundamental requirement for quantization,\cite{IT18JCP,1987RISBPhysRevB.36.2770,IT19JCP,Lipeng3Drotar2019}  while the CL-QFPE coincides with the classical one ($i.e.,$ $\hbar$ does not appear).\cite{ST02JPSJ,ST03JCP}  

To perform numerical calculation, we set $\hbar = 1$, $q = -1$, $m_{S} = 0.5$, and $r_0 = 1$, and thus we have $\omega_0 = 1.0$.  The magnetic flux is expressed in units of the flux quanta, $\bar\Phi = \Phi/\Phi_0$, where $\Phi_0 = h / q$.  We consider two cases with (a) $\bar\Phi  = 0$ and (b) $\bar\Phi= 1$.  
The grid size of the WDF in the momentum directions are set to $| p_n |/\hbar  \leq 31 / 2$ in the RISB case and $N_{\rm p} = 128$ in the CL case with the grid size $d p = 0.25$, whereas in the coordinate directions, the grid size is set to $N_\theta = 64$ in both cases, with the range $\theta \in [ 0, 2 \pi )$.
A second-order central difference scheme was adopted as the expression for the derivatives of coordinates and momentum. For the numerical time integration, the implicit method, which solves the partial differential equation accurately and simultaneously, was used with the periodic boundary condition $W(p_n, \theta) =W(p_n, \theta + 2 \pi) $. 

\begin{figure}
    \centering
    \includegraphics[width=1\linewidth]{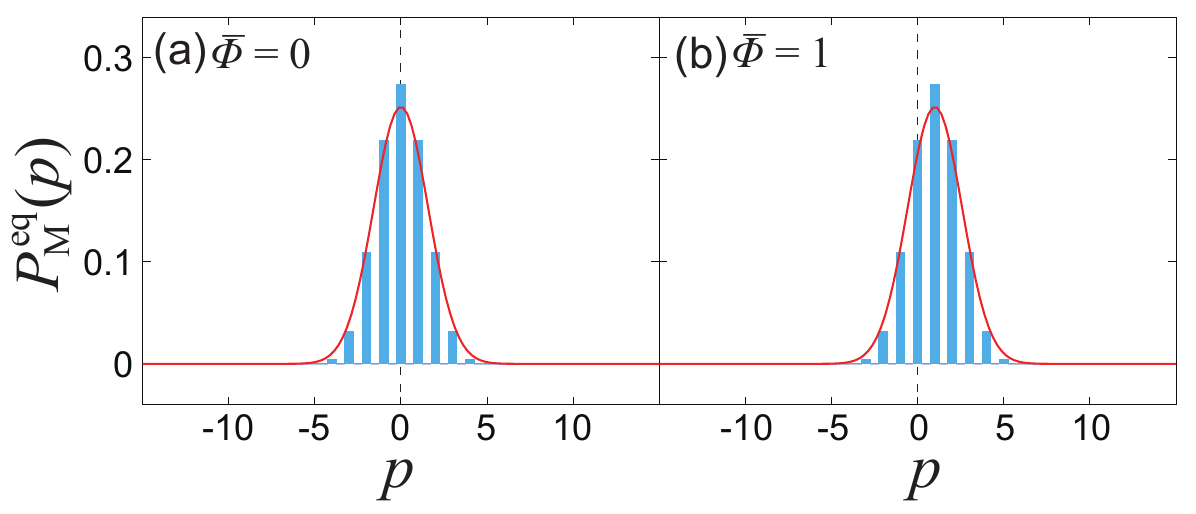}
    \caption{The equilibrium distribution in the momentum space for the single charge flow obtained from the RISB model [Eq. \eqref{eqn:FP_RISB}] (blue bars) and the CL model [Eq. \eqref{eqn:FP_CL}] (red curves). Here we set $\eta = 0.01$, $\beta = 0.2$ for magnetic flux (a) $\bar\Phi  = 0$ and (b) $\bar\Phi= 1$ for $\bar\Phi\equiv \Phi/\Phi_0$. In the case of the RISB model, unlike the CL model, which exhibits a continuous distribution, it shows a discretized distribution that exists only at integer values.  However, in both cases, the distribution follows a Gaussian profile centered at $\bar\Phi$. }
    \label{fig:St_St}
\end{figure}

In Fig. \ref{fig:St_St}, we present the equilibrium distribution in the momentum space obtained from Eqs.\eqref{eqn:FP_RISB} and \eqref{eqn:FP_CL} for different magnetic fluxes.
 The distribution function is defined as follows: $P_{\rm CL}^{\rm eq} ( p ) = \int_0^{2 \pi} W_{\rm CL} ( p , \theta ) d \theta$ for the CL model, and $P_{\rm RISB}^{\rm eq} ( p_n ) = ({1}/{2}) \int^{2 \pi}_0 W ( p_n , \theta ) d \theta$ for the RISB model.  
In the CL case, $P_{\rm CL}^{\rm eq} ( p )$  exhibits the Gaussian profiles centered at $p = q r_0 A$.
This can be readily discerned from the steady-state solution of Eq. \eqref{eqn:FP_CL} expressed as 
\begin{align}
P_{\rm CL}^{\rm eq} ( p ) = \frac{1}{\mathcal{N}} \exp \left[ - \frac{\beta}{2 I_{S}} ( p - q r_0 A )^2 \right] ,
\label{eqn:WigRISB}
\end{align} 
where  $\mathcal{N}$ is the normalization constant.  In the case of RISB, the energy state is discrete due to the DWT-PBC. However, the outline profile is also a Gaussian centered at
$p = q r_0 A$. Due to the coupling between radial and rotational degrees of freedom, fluctuations are suppressed, resulting in a narrower momentum distribution in the RISB outcome compared to the CL result.

Note that the Markov approximation requires a high-temperature limit for the bath, resulting in a very wide Gaussian distribution in both CL and RISB cases. This effectively suppresses the AB shift at high temperatures. We should also mention that the observed shift in both CL and RISB cases can be attributable to the gauge invariance of their equations of motion, and the shifts themselves are not necessarily quantum effects: The CL-QFPE has the same form even in the classical limit, showing that a shift exists even in the classical case.

\subsection{Linear response: Markovian case}

In the absence of the gauge field, the linear absorption spectrum of 2D and 3D damped free rotors has been investigated through the use of the Markovian QME, employing the eigenstate representation.\cite{IT18JCP,IT19JCP} 
Here, we calculate the linear response spectrum of the AB ring system for laser excitation in accordance with the DWT-PBC formalism. 
Although there is no permanent dipole in the AB ring, the transient dipole allows for measurement of the spectrum, as in the case of the 2D damped free rotor.\cite{ST02JPSJ,ST03JCP,IT18JCP}  
The linear repose (rotational) spectrum is expressed as
\begin{align}
\sigma (\omega) = {\rm Im} \left\{\int_0^{\infty} dt \ {\rm e}^{i \omega t }
{R}^{(1)}(t) 
\right\}, \label{eqn:ProjCurr}
\end{align}
where the first-order response function is defined as
${R}^{(1)}(t) = {i}\langle [\cos{\hat\theta(t )},  \cos{\hat\theta (0)} ] \rangle/{\hbar}$.
In the Markovian case, we can rewrite the response function as\cite{T06JPSJ, IT19JCP}
\begin{eqnarray}
\label{R_1t}
R^{(1)}(t) = \frac{i}{\hbar}{\rm Tr}\left\{ \cos\hat\theta \hat{\mathcal{G}} (t) 
( \cos\hat\theta )^\times W^{\rm eq} \right\},
\end{eqnarray}
where $\hat{\mathcal{G}}(t)$ is Green's function in the absence of a laser interaction, and $W^{\rm eq}$ is the equilibrium WDF.

In the CL case, $W_{\rm CL}^{\rm eq}$ is obtained as the steady-state solution of Eq.\eqref{eqn:FP_CL}. After laser excitation, the state of the system becomes $\cos\hat\theta^\times W^{\rm eq}$. Using that as the initial condition, integrate Eq \eqref{eqn:FP_CL} to $t$ to obtain $W' _{\rm CL}( t ) = \hat{\mathcal{G}} ( t ) ( \cos \hat{\theta} )^\times W_{\rm CL}^{\rm eq}$.  Then $R^{(1)}(t)$ is evaluated as the expectation value $\langle \cos{\hat\theta(t )} \rangle = {\rm Tr} \{ \cos{\hat\theta} W'_{\rm CL} ( t )  \}$ for the perturbed distribution $W'_{\rm CL}(t)$.\cite{T06JPSJ,IT18JCP}

In the RISB case, Eq.\eqref{R_1t} is expressed in the DWT-PBC formulation as
\begin{align}
\label{eq:LinearResponseRISB}
& R^{(1)} ( t ) = \frac{1}{2} \sum_{n , l = - \infty}^{+ \infty} 
\int_0^{2 \pi} d \theta d \theta' \cos \theta \; \mathcal{G} ( \theta , p_n ; \theta' , p_l ; t ) \nonumber \\ 
&~~~~~~~~~ \times \Big[ \sin \theta' \big( W^{\rm eq} ( p_{l + 1} , \theta' ) - W^{\rm eq} ( p_{l - 1} , \theta' ) \big) \Big] ,
\end{align}
where $\mathcal{G} ( \theta , p_n ; \theta' , p_l ; t )$ is the Green's function in the DWT-PBC representation.
The evaluation of $W^{\rm eq} ( p_n , \theta )$ and $\mathcal{G} ( \theta , p_n ; \theta' , p_l ; t )$
are now performed by numerical integration of Eq. \eqref{eqn:FP_RISB}. In Eq.~\eqref{eq:LinearResponseRISB}, when we aply the first pulse to the system, we compute $W' ( \theta , p_n ; t = 0 ) = \sin \theta ( \delta / \delta p_n ) W ( p_n , \theta )$. Then, we compute the time evolution for the WDF, $W' ( p_n , \theta )$ using Eq.~\eqref{eqn:FP_RISB}.

\begin{figure}
    \centering
    \includegraphics[width=0.75\linewidth]{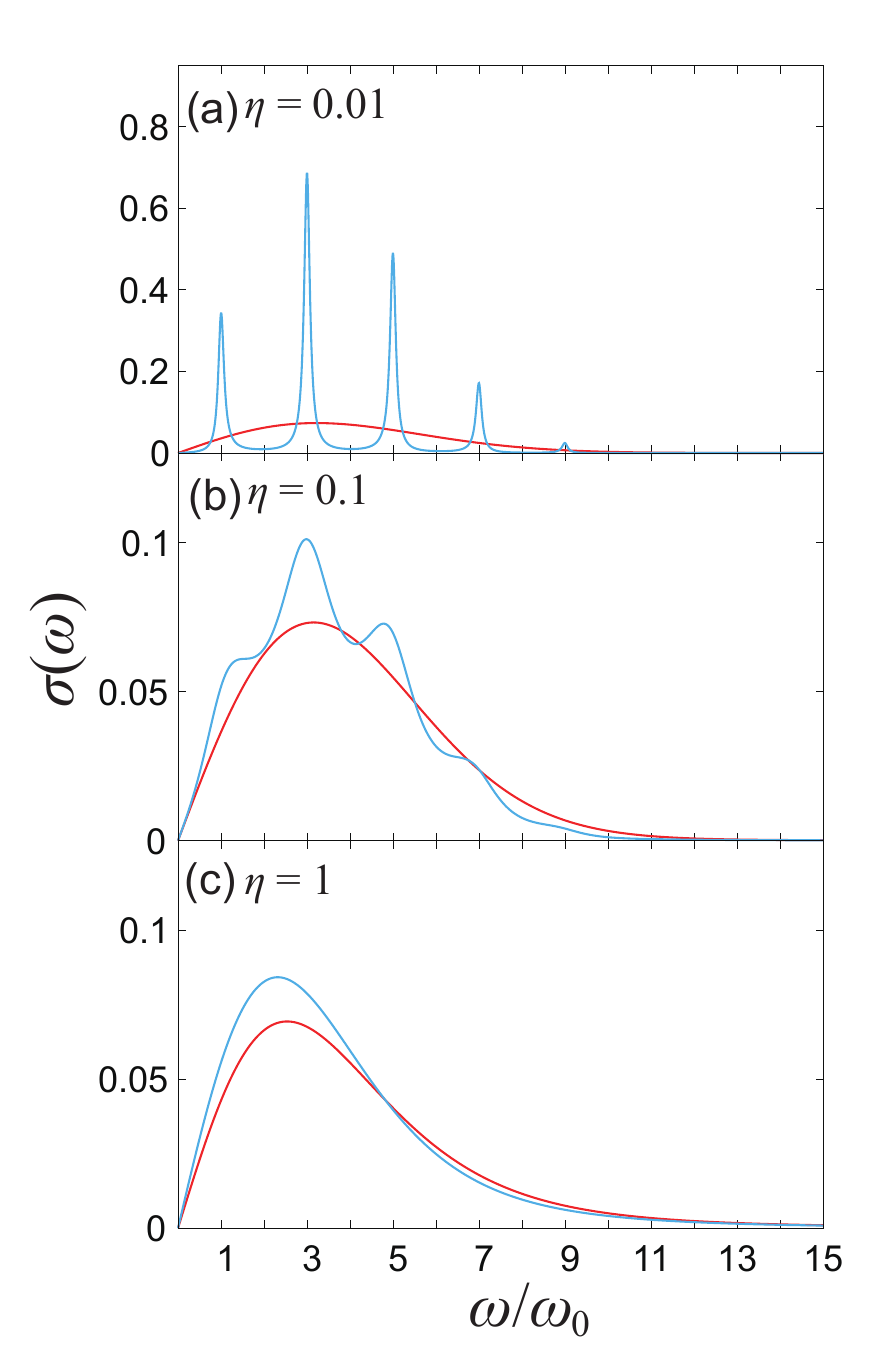}
    \caption{Rotational spectrum without magnetic flux ($i.e. \bar\Phi=0$) are shown for the RISB (blue curves) and CL (red curves) cases under fixed $\beta = 0.2$, with (a) $\eta = 0.01$, (b) $  0.1$,  and (c) $  1.0$ for fixed $\beta = 0.2$.} 
    \label{fig:ProjCurr}
\end{figure}

To verify the accuracy of the description of the Wigner function based on the DWR-PBC, we first considered the case where the magnetic flux is zero.  In Fig. \ref{fig:ProjCurr} we depict the rotational spectrum for different S-B coupling strengths. These results are consistent with the absorption spectrum of a damped free rotor obtained previously,\cite{IT18JCP,IT19JCP} demonstrating the validity of our U(1)-QHFP formulation. In this case, rotational bands correspond to the transitions between different angular momentum states. They are observed in the RISB model in the case of weak S-B coupling,\cite{IT18JCP,IT19JCP}  while a single featureless broadened peak is observed in the CL model due to the lack of rotational symmetry.\cite{ST02JPSJ,ST03JCP}   As the S-B coupling strength increases, the quantum coherence of the RISB model is suppressed, leading to the disappearance of sharp peaks. However, the remaining peak in the RISB model undergoes a red shift due to copling between the rotational and radial degrees of freedom.

\begin{figure}
    \centering
    \includegraphics[width=1\linewidth]{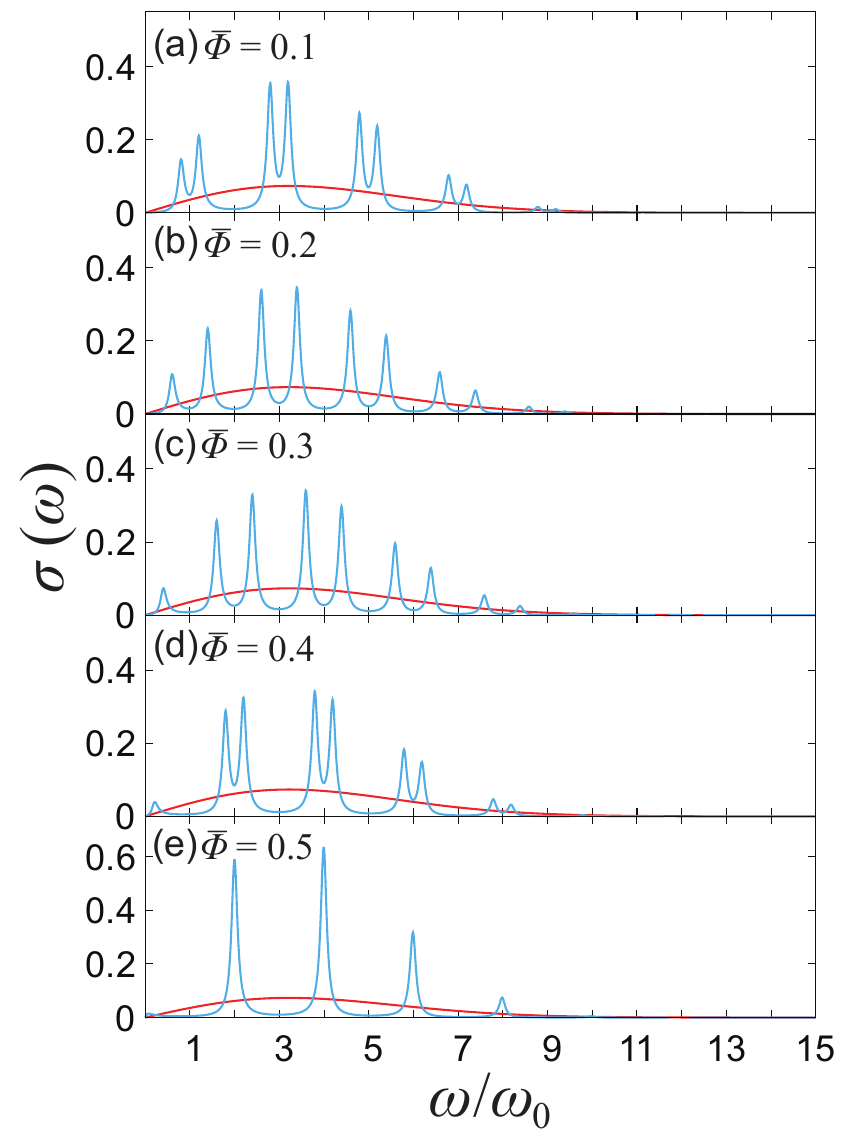}
    \caption{Rotational spectrum under the influence of the vector potential. Each spectrum is obtained with $\eta = 0.01$ and $\beta = 0.2$ 
    at various magnetic fluxes: (a) $\bar\Phi = 0.1$, (b) $0.2$, (c) $0.3$, (d) $0.4$, and (e) $0.5$. The case for $\bar\Phi = 0$ is presented in Fig. \ref{fig:ProjCurr} (a).}
    \label{fig:ProjCurr_A}
\end{figure}

We now discuss the effects of a magnetic flux. Figure  \ref{fig:ProjCurr_A} depicts the rotational spectrum for various values of the flux strengths: (a) $\bar\Phi=0.1$, (b) $0.2$, (c) $0.3$, (d) $0.4$, and (e) $0.5$ with fixed $\eta = 0.01$ and $\beta=0.2$.  The results for $\bar\Phi = 0.0$ is presented in Fig. \ref{fig:ProjCurr} (a).
As the value of $\bar\Phi$ increases, each peak (rotational band) observed at $\bar\Phi=0$ in Fig. \ref{fig:ProjCurr} (a) splits into two as depicted in Fig. \ref{fig:ProjCurr_A}(b)-(d). The distance between two peaks increases as $\bar\Phi$ increases up to around $\bar\Phi=0.25$, and then begins to decrease, disappearing at (e) $\bar\Phi=0.5$.

This phenomenon can be explained by analyzing the transition energy of the AB ring in the absence of the bath: The eigenfunction of the system Hamiltonian is given by $\psi_n ( \theta ) = e^{i n \theta} / \sqrt{2 \pi}$, where the corresponding eigenenergy is expressed as
\begin{align}
\label{eq:SystemEigenenergy}
E_n ( \Phi ) = \hbar \omega_0 \left( n - \bar{\Phi} \right)^2 .
\end{align}
Then the transition energy from state $\pm |n|$ to $\pm ( |n| + 1 )$ is evaluated as $\Delta E_{\pm |n| \rightarrow \pm (|n| + 1)} = E_{\pm (|n| + 1)} - E_{\pm |n|} = \hbar \omega_0 ( 2 | n | + 1 \mp 2 \bar{\Phi} )$.\cite{ByersYang1961}  For the cases shown in Figs. \ref{fig:ProjCurr} and \ref{fig:ProjCurr_A}, this is expressed as $(2 n + 1) \pm 2 \bar{\Phi}$, where $n$ is a non-negative integer.

Thus, in the case of Fig. \ref{fig:ProjCurr}, where $\bar\Phi = 0$, rotational spectrum exhibits peaks exclusively at positions corresponding to $(2n + 1)$, reflecting transitions between angular momentum eigenstates. 
In Fig. \ref{fig:ProjCurr_A}, when $\bar\Phi$ becomes finite and smaller than $\bar\Phi<0.5$, each rotational band split because the transition now becomes  $( 2 n + 1) \pm 2 \bar\Phi$.   In Fig. \ref{fig:ProjCurr_A} (e) $\bar\Phi=0.5$, the splits disappear because for $\bar\Phi=\pm 0.5$, the energy spacing becomes $\Delta E = (2n + 1) \pm 1 $, leading to absorption peaks located at the positions, $2n$ excluding zero. Because $\{\bar\Phi \in [0, 0.5) \mid {\rm mod}\ 0.5\}$, the rotational bands split again for $0.5 \le \bar\Phi <1 $ as Fig. \ref{fig:ProjCurr_A} (e) $\rightarrow$ Fig. \ref{fig:ProjCurr_A} (a) as $\bar\Phi $ increases.

\subsection{Persistent Current: Non-Markovian cases}
\label{sec:ABcurrent}

Persistent current is a unique property exhibited by AB rings.\cite{1992FukuyamaAndo,BookImry,Fomin2018}
 It has been observed not only in superconducting materials\cite{ByersYang1961,1986HalperinRevModPhys.58.533,1991ii,Blasi_ACring2023} but also in metals and semiconductors,\cite{1986ABPhysRevLett.56.386,1987Umbach10.1063/1.97887,levy1990magnetization,bluhm2009persistent,PC2010}
 where the ring circumference is sufficiently short and the phase of the electron wave function is preserved.
As shown in the previous subsection, the gauge effects that appear in rotational spectrum exist even in the presence of a thermal bath. Here, we investigate the environmental effects on persistent current using the CL and RISB models.
The important point here is that persistent current manifests exclusively at low temperatures, where the coherence of the wave function can be sustained. Therefore, here we utilize the U(1)-QHFPE, in which the temperature restrictions have been removed. 

The current contribution from the state $n$ is given by the Byers--Yang relation expressed as\cite{ByersYang1961,levy1990magnetization} 
\begin{eqnarray}
I_n ( \Phi ) = - \frac{\partial E_n ( \Phi )}{\partial \Phi} = \frac{2 \hbar \omega_0}{\Phi_0}
( n - \bar{\Phi} ),
\label{eqn:Current}
\end{eqnarray}
where the system energy is given by Eq. \eqref{eq:SystemEigenenergy}. 
In natural units, this upper bound coincides precisely with the dimensionless magnetic flux $\bar\Phi$, which serves as a distinctive signature of the AB effect. To substantiate this feature within our framework, we evaluated the expectation value of the current. In equilibrium, this quantity can be formulated via the momentum distribution function, $P_{\rm RISB}^{\rm eq} ( p_n )$, as  $\hbar \sum_n I_n ( \Phi ) P_{\rm RISB}^{\rm eq} ( p_n )$. This expression enables a direct assessment of the current based on the WDF as follows:
\begin{eqnarray}
J^{\rm eq} ( \bar \Phi ) &&\equiv {\rm Tr} \left\{ \hat{J} \hat{\rho}^{\rm eq} \right\} \nonumber \\
&&= \frac{\hbar}{2} \sum_n \int_0^{2 \pi} d \theta 
\left[ \frac{2 \omega_0}{\Phi_0} ( p_n - \hbar \bar{\Phi} ) W_{\{ \bm{0} \}}^{\rm eq} ( p_n , \theta) \right] ,\nonumber \\
\label{eqn:Current2}
\end{eqnarray}
where $\hat{J} \equiv 2 \omega_0 ( \hat{p}_\theta - q r_0 A) / \Phi_0$ is the current operator.

For the computation of the persistent current, we set the inverse temperature as (a) $\beta = 2.5$ (low) and (b) $\beta = 0.2$ (high).  For the U(1)-QFHPE in Eq.~\eqref{eq:QHFPE-C}, we set the truncation number of the hierarchy to $N =  2$ for the very weak S--B coupling case  ($\eta = 1.0 \times 10^{-3}$) and $N = 8$ for the strong coupling case ($\eta = 1.0$).
The number of Pad{\'e} frequencies are chosen to $K = 1$ and $4$ at high ($\beta = 0.2$) and low ($\beta = 2.5$) temperatures, respectively. The grid size of the WDF was set to $N_\theta = 64$ and $| p_n |/\hbar  \leq 31 / 2$. Numerical integration for the U(1)-QHFPE was performed using the Runge--Kutta--Fehlberg method.\cite{fehlberg1969low} 

\begin{figure}
    \centering
    \includegraphics[width=1\linewidth]{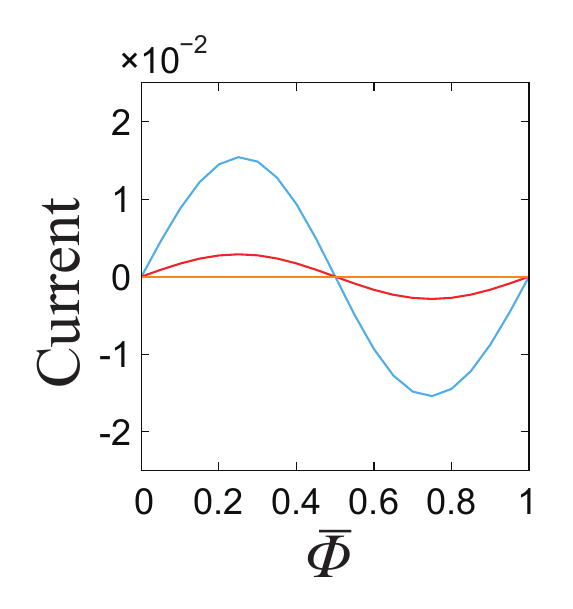}
    \caption{Persistent current computed from the U(1)-QHFPE is plotted as a function of the magnetic flux $\bar{\Phi}$ for very weak ($\eta = 1.0 \times 10^{-3}$, blue curves) and the strong ($\eta = 1.0$, red curves) S-B coupling cases at the low temperature $\beta = 2.5$. 
The orange line represents the high-temperature case ($\beta = 0.2$) with the weak S-B coupling ($\eta = 1.0 \times 10^{-3}$): The persistent current vanishes at this high-temperature regime, regardless of the coupling strength, aligning with the behavior of the orange line. Similarly, Markovian results—relying on high-temperature approximations—also show no persistent current.(not shown.) }
    \label{fig:non-Markovian_Current}
\end{figure}

In Fig.~\ref{fig:non-Markovian_Current}, we plot the current as a function of the magnetic flux $\bar{\Phi}$. The blue and red curves represent the results in the very weak ($\eta = 1.0 \times 10^{-3}$) and strong ($\eta = 1.0$) S-B coupling cases at the low temperature ($\beta = 2.5$), respectively, while the orange line represents the result in the high temperature case ($\beta = 0.2$) with the very weak S-B coupling ($\eta = 1.0 \times 10^{-3}$).   

At low temperatures, the system remains in its ground state. From Eq. \eqref{eqn:Current}, when the magnetic flux $\bar{\Phi}$ is small, the current increases proportionally to the flux, as $I_0(\Phi) \propto \bar{\Phi}$. However, as $\bar{\Phi}$ increases, the eigenenergy of the first excited state ($n=-1$) decreases, following $\Delta E_{0 \rightarrow -1} = \hbar \omega_0 (1 - 2 \bar{\Phi})$, and becomes thermally populated.  Then, for $0.25 < \bar{\Phi}$, the current associated with the first excited state flows in the opposite direction to the ground state current according to Eq. \eqref{eqn:Current}. As a result, the total current decreases with increasing $\bar{\Phi}$. When $\bar{\Phi} = 0.5$, the current vanishes because the ground and first excited states become degenerate, canceling the net current. For $0.5 \leq \bar{\Phi} < 1$, the roles of $n=0$ and $n=-1$ switch, causing the current to flow in the opposite direction.

As the S-B coupling strength increases (red curves), higher excited states become more populated, resulting in a further decrease in the net current.
At high temperatures (orange line), the current is completely suppressed because the eigenstates responsible for the rightward and leftward currents achieve equal population.

The periodicity of the current can also be demonstrated using the U(1)-QHFPE framework [Eq.~\eqref{eq:QHFPE-C}]. When the magnetic flux changes as $\Phi \rightarrow \Phi + k \Phi_0$, where $k$ is an integer, Eq.~\eqref{eq:QHFPE-C} remains invariant under the transformation $W_{\{ \bm{n}_\alpha \}} ( p_n , \theta ) \rightarrow W_{\{ \bm{n}_\alpha \}} ( p_{n - 2 k} , \theta )$. Because the thermal equilibrium state is unique, we have $W^{\rm eq}_{\{ \bm{n}_\alpha \}} ( p_n , \theta) \big|_{\bar \Phi}  =W^{\rm eq}_{\{ \bm{n}_\alpha \}} ( p_{n + 2 k} , \theta ) \big|_{\bar \Phi + k} $
for the equilibrium state. 
Thus, from Eq.~\eqref{eqn:Current2}, we find the periodicity of persistent current  expressed as $J^{\rm eq} ( \Phi + k \Phi_0 ) = J^{\rm eq} ( \Phi )$.
In the weak-coupling limit, this result aligns with the Byers--Yang theorem.\cite{ByersYang1961}

\section{Conclusion}
\label{sec:Conclution} 

We considered a system with a gauge field that interacts with multiple heat baths (the RISB model).  Assuming Drude SDF for each bath and by reducing the bath degrees of freedom, we derived the U(1)-QHFPE, which can treat anisotropic S-B interaction in a non-perturbative and non-Markovian manner. For comparison, we also considered a CL model including a gauge field, and derived the CL U(1)-QHFPE. Both of these equations satisfy gauge invariance.

To verify the validity of the equations of motion, we consider the AB ring system and investigate the equilibrium distribution, rotational spectrum in the Markov case, and the persistent current in the non-Markovian case. While the RISB models exhibited rotational bands reflecting discrete energy levels, the CL models showed only a broad peak due to their lack of rotational symmetry.\cite{IT18JCP} Under a gauge field, the peaks in the RISB model split due to magnetic flux-induced transitions, while the change in the CL model is minor.

The flow of the persistent current depends on the thermal distribution of the AB ring in momentum space. At high temperatures, the effect of the gauge field is suppressed due to the broadening of the equilibrium distribution. In such cases, the momentum shift caused by the gauge field has only a minor effect. Moreover, since actual experiments are performed at low temperatures, methods such as the U(1)-HEOM and U(1)-QHFPE are suitable for studying the persistent current.

Not only in the case of RISB but also in the case of CL, when the temperature was high, no persistent current. was observed. Thus, the Markovian approximation should not be
employed for the investigation of the persistent current.

From this point onward, the CL model becomes appropriate for describing rings whose size exceeds the electronic coherence length,\cite{TW92JCP,KT13JPCB,IDT19JCP} whereas the RISB model remains suitable for nano-rings with sub-coherence dimensions. Notably, in the presence of an isotropic thermal bath, enlarging the ring size leads the RISB model to asymptotically converge to the CL description.

While a system in thermal equilibrium, defined by the inverse temperature $\beta$, means that the system is not generating heat or entropy. Using the present formalism based on HEOM, it is possible to evaluate thermodynamic quantities such as heat and entropy not only in equilibrium systems but also in non-equilibrium systems.\cite{KT24JCP1,KT24JCP2}

The equations of motion derived in this paper allow for the introduction of arbitrary time-dependent external fields. This makes it possible to study the AB effect under time-dependent external fields. 
The source code used in this paper is provided in a separate paper.\cite{KYT25JCP2} 

As demonstrated herein, the core principle of U(1)-HEOM under gauge and rotational invariance resides in constructing the total Hamiltonian, including the thermal bath, in a manner that rigorously preserves each symmetry. Accordingly, equivalent results are anticipated from 
the quasi-adiabatic path-integral (QUAPI) approach\cite{Makri95, Makri96, Makri96B} 
and pseudomode (PM) approaches,\cite{Burghardt2011,
Plenio2018,Lambert2019,Petruccione2020,pseudomodesNori2024} provided that the bath degrees of freedom are eliminated in a symmetry-consistent fashion. While the simulation of more realistic scenarios demands computational resources that exceed the capabilities of current classical hardware, quantum computers may ultimately be required to perform such calculations~\cite{Dan2024qHEOM,Nori2024PRR,Shi2025MCEHEOM}.

Advancing research in the aforementioned directions remains a compelling challenge for the future.


\section*{Acknowledgments}
Y. T.  thanks Kensuke Kobayashi for the provision of valuable experimental and theoretical information concerning the AB ring.
Y. T. was supported by JST (Grant No. CREST 1002405000170). H. Y. acknowledges a fellowship supported by JST SPRING, the establishment of university fellowships toward the creation of science technology innovation (Grant No.~JPMJSP2110). S. K.  was supported by Grant-in-Aid for JSPS Fellows (Grant No.24KJ1373).

\section*{Author declarations}
\subsection*{Conflict of Interest}
The authors have no conflicts to disclose.

\section*{Data availability}
The data that support the findings of this study are available from the corresponding author upon reasonable request.

\appendix

\section{Gauge invariance of the U(1)-HEOM}
\label{Sec:GIHEOM}
Under this transformation, Eqs.  \eqref{eq:HEOM}-\eqref{eq:Theta_0}, is expressed as
\begin{eqnarray}
\label{eqn:gaugeHEOM}
\frac{\partial}{\partial t} \hat{\rho}'_{\{ \bm{n}_\alpha \}} ( t )
&&= - \left[ \frac{i}{\hbar} {\hat H}_{S}^{\prime \times} ( t ) 
+ \sum_\alpha^{x , y , z} \sum_{j = 0}^{K_\alpha} n_j^\alpha \nu_j^\alpha \right] \!
\hat{\rho}'_{\{ \bm{n}_\alpha \}} ( t ) \nonumber \\
&& + \sum_\alpha^{x , y , z}   \sum_{j = 0}^{K_\alpha}
\hat{\Phi}^\alpha  \hat{\rho}'_{\{ \bm{n}_\alpha + \bm{e}^j_\alpha \}} ( t )\nonumber \\
&&
+ \sum_\alpha^{x , y , z} \sum_{j = 0}^{K_\alpha}
n_j^\alpha \hat{\Theta}_j^{\prime \alpha} 
\hat{\rho}'_{\{ \bm{n}_\alpha - \bm{e}^j_\alpha \}} ( t ),
\end{eqnarray}
where 
\begin{align}
\hat{\rho}'_{\{ \bm{n}_\alpha \}} = U ( \hat{\bm{x}} , t ) \hat{\rho}_{\{ \bm{n}_\alpha \}} ( t )
U^\dagger ( \hat{\bm{x}} , t ) 
\end{align}
and ${\hat H}'_{S} ( t )$ and $\hat{\Theta}_0^{\prime \alpha}$ are the Hamiltonian and relaxation operator after gauge transformation, $\bm{A}' ( \bm{x} ; t)$ and $\phi' ( \bm{x} ; t)$, respectively. The gauge transformation directly affects the operator $\hat{\Theta}_0^\alpha ( t )$, while $\hat{\Theta}_j^\alpha \: ( j \neq 0 )$ is invariant under gauge transformations.
Using Eq. \eqref{eqn:U1}, we have
\begin{align}
\hat{H}'_{S} ( t ) &= U ( \hat{\bm{x}} , t ) \hat{H}_{S} ( t ) \hat{U}^\dagger ( \hat{\bm{x}} , t ) - i q \frac{\partial \chi( \hat{\bm{x}} , t )}{\partial t}  \nonumber \\
&= \frac{1}{2m_{S}} \left[ \hat{\bm{p}} - q \bm{A} ( \hat{\bm{x}} ; t) 
- q \nabla \chi ( \hat{\bm{x}} ; t ) \right]^2 \nonumber \\
&+ \phi ( \hat{\bm{x}} ; t ) - q \frac{\partial \chi ( \hat{\bm{x}} , t )}{\partial t} ,
\end{align}
and 
\begin{align}
\hat{\Theta}_0^{\prime \alpha} ( t )
&
= - \frac{i}{\hbar} \bigg[ \frac{\eta_\alpha \gamma_\alpha}{\beta} 
\left( 1 + \sum_{j = 1}^{K_\alpha} 
\frac{2 \bar{\eta}_j^\alpha \gamma_\alpha^2}{\gamma_\alpha^2 - ( \nu_j^\alpha )^2} \right)
 \nonumber \\
&\times \left( U ( \hat{\bm{x}} , t ) \hat{V}_\alpha U^\dagger ( \hat{\bm{x}} , t ) \right)^\times \nonumber \\
&
\quad \qquad
+ \frac{i \hbar \eta_\alpha \gamma_\alpha}{2} \left( U ( \hat{\bm{x}} , t ) \hat{\dot{V}}_\alpha
U^\dagger ( \hat{\bm{x}} , t ) \right)^\circ \bigg] \nonumber \\
&
= - \frac{i}{\hbar} \left[ \frac{\eta_\alpha \gamma_\alpha}{\beta} 
\left( 1 + \sum_{j = 1}^{K_\alpha} 
\frac{2 \bar{\eta}_j^\alpha \gamma_\alpha^2}{\gamma_\alpha^2 - ( \nu_j^\alpha )^2} \right) 
\hat{V}_\alpha^\times  \right. \nonumber \\
&
\left. + \frac{i \hbar \eta_\alpha \gamma_\alpha}{2} \hat{\dot{V}}_\alpha^{\prime \circ} ( t ) \right] ,
\end{align}
where $\hat{\dot{V}}'_\alpha ( t ) = i [ \hat{H}'_{S} ( t ) , \hat{V}_\alpha ] / \hbar$.
Based on the above results, Eq.~\eqref{eqn:gaugeHEOM} reduces to Eq.~\eqref{eq:HEOM} again, and the U(1)-HEOM satisfies gauge invariance.

\section{U(1)-HEOM for the AB ring} \label{QHFPE_ABring}

To apply the U(1)-HEOM to an AB ring, we consider the 2D case and assume the isotropic environment $\eta_x = \eta_y \equiv \eta$ and $\gamma_x = \gamma_y \equiv \gamma$ in Eq.~\eqref{JDrude}. To reduce the radial degree of freedom later, we incorporate the counter term to the system Hamiltonian expressed as
\begin{eqnarray}
\label{eq:SystemHamiltonianC}
\hat{H}_{S}^{C}  = \hat{H}_{S} + \frac{\eta \gamma}{2} 
( \hat{q}_x^2 + \hat{q}_y^2 ) .
\end{eqnarray}
Then, we can express the U(1)-HEOM as
\begin{eqnarray}
\label{eq:HEOM-C}
\frac{\partial \rho_{\{ \bm{n}_\alpha \}}^{C}   ( \bm{q} , \bm{q}' ) }{\partial t} 
&&=  - \left[ \frac{i}{\hbar} \hat{H}_{S}^{{C}  \times} 
+ \sum_{\alpha}^{x , y} \sum_{k = 0}^K n_k^\alpha \nu_k \right]
\rho^{C} _{\{ \bm{n}_\alpha \}} ( \bm{q} , \bm{q}' )\nonumber \\
&&+ \sum_\alpha^{x , y} \sum_{k = 0}^K 
\hat{\Phi}^\alpha \rho_{\{ \bm{n}_\alpha + \bm{e}_\alpha^k \}}^{C}  ( \bm{q} , \bm{q}' ) \nonumber \\
&&
+ \sum_\alpha^{x , y} \sum_{k = 0}^K n_k^\alpha \hat{\Theta}_k^{{C}  , \alpha}
\rho_{\{ \bm{n}_\alpha - \bm{e}_\alpha^k \}}^{C}  ( \bm{q} , \bm{q}' ) ,
\end{eqnarray}
where $\hat{\rho}_{\{ \bm{n}_\alpha \}}^{C} $ is an auxiliary density operator in which the counter term is incorporated into the system Hamiltonian, and we define the operators as
\begin{align}
\label{eq:defOperatorC1x}
\hat{\Theta}_{\alpha , 0}^{C}  = 
- \frac{i}{\hbar} c_0 ( q_\alpha - q'_\alpha )
- \frac{\eta \gamma^2}{2} ( q_\alpha + q'_\alpha ) ,
\end{align}
and $\hat{\Theta}_{\alpha , k}^{C}  = \hat{\Theta}_k^\alpha$ for $k = 1 , 2 , \cdots , K$. Here, we set $c_0 = \frac{\eta \gamma}{\beta} \left( 1 + \sum_{j = 1}^K \frac{2 \bar{\eta}_j \gamma^2}{\gamma^2 - \nu_j^2} \right)$. In Eq.~\eqref{eq:HEOM-C}, we also set $K_x = K_y \equiv K$. 

Using the polar coordinate, we can express the Cartesian coordinate variables as $q_x = r \cos \theta$ and $q_y = r \sin \theta$ and express the auxiliary density operator (ADO) as $\rho_{\{ \vec{n}_\alpha \}}^{C}  ( r, \theta, r', \theta'; t )$. 
We reduce the radial degree of freedom for $r_0$.  The reduced system density operator is then expressed as
\begin{align}
\label{eq:defReducedDensityOperator}
R_{\{ \bm{n}_\alpha \}}^{C}  ( \theta , \theta' ; t ) = \int_0^\infty r 
\rho_{\{ \bm{n}_\alpha \}}^{C}  ( r , \theta , r' , \theta' ; t ) d r .
\end{align}
For a fixed radius, the following equality holds for an arbitrary real number $c$.
\begin{align}
\label{eq:defReducedDensityOperator2}
r_0^c R_{\{ \bm{n}_\alpha \}}^{C}  ( \theta , \theta' ; t ) = \int_0^\infty r^{c + 1}
\rho_{\{ \bm{n}_\alpha \}}^{C}  ( r , \theta , r' , \theta' ; t ) d r .
\end{align}
Then, we obtain the U(1)-HEOM for the AB ring system as follows:
\begin{eqnarray}
\label{eq:HEOM-C2}
\frac{\partial R_{\{ \bm{n}_\alpha \}}^{C}   ( \theta , \theta' )}{\partial t}  &&
= - \left[ \frac{i}{\hbar} \hat{H}_{S}^\times 
+ \sum_\alpha^{x , y} \sum_{k = 0}^K n_k^\alpha \nu_j \right]
R_{\{ \bm{n}_\alpha \}}^{C}  ( \theta , \theta' ) \nonumber \\
&&+ \sum_\alpha^{x , y} \sum_{k = 0}^K \hat{\Phi}_{\rm red}^\alpha 
R_{\{ \bm{n}_\alpha + \bm{e}_\alpha^k \}}^{C}  ( \theta , \theta' ) \nonumber \\
&&
+ \sum_\alpha^{x , y} \sum_{k = 0}^K n_k^\alpha \hat{\Theta}_{\alpha , k}^{\rm red, C}
R_{\{ \bm{n}_\alpha - \bm{e}_\alpha^k \}}^{\rm red , C} ( \theta , \theta' ) ,
\end{eqnarray}
where $\hat{\Phi}_{\rm red}^\alpha$ and $\hat{\Theta}_{\alpha , k}^{\rm red, C}$ are the operators $\hat{\Phi}^\alpha$ and $\hat{\Theta}_{\alpha , k}^{C} $ in which $q_x$, $q_y$, $q'_x$, and $q'_y$ are replaced by $r_0 \cos \theta$, $r_0 \sin \theta$, $r_0 \cos \theta' $, and $r_0 \sin \theta'$, respectively. Performing the Wigner transformation to Eq.~\eqref{eq:HEOM-C2}, we obtain the U(1)-Quantum Hierarchical Fokker--Planck Equations [U(1)-QHFPE] in Eq.~\eqref{eq:QHFPE-C}.\\

\section{U(1)-QFPE for the AB ring}
\label{sec:QFPE-ABring}
When the QME [Eq.~\eqref{eq:MasterEq}] is applied to the AB ring system, it can be reformulated in polar coordinates as follows:
\begin{widetext}
\begin{align}
\frac{\partial}{\partial t} \rho_0 ( r , \theta , r' , \theta' ; t ) 
&
= \left\{ - \frac{i}{\hbar} \hat{H}_S^\times
 - \frac{\eta}{\beta \hbar^2}
\left( r^2 + r^{\prime \; 2} - 2 r r' \cos ( \theta - \theta' ) \right) \right\} 
\rho_0 ( r , \theta , r' , \theta' ; t ) \nonumber \\
&
~~~~ - \frac{\eta}{2 m_S} 
\bigg[ ( r - r' \cos ( \theta - \theta' ) ) \frac{\partial}{\partial r}
- ( r' - r \cos ( \theta - \theta' ) ) \frac{\partial}{\partial r'} \bigg] \rho_0 ( r , \theta , r' , \theta' ; t ) 
\nonumber \\
&
~~~~ - \frac{i \eta}{2 m_S \hbar}
\left[ \frac{r'}{r} \sin ( \theta - \theta' ) \left( - i \hbar \frac{\partial}{\partial \theta} - q r A_\theta \right)
+ \frac{r}{r'} \sin ( \theta - \theta' ) \left( i \hbar \frac{\partial}{\partial \theta'} - q r' A_\theta \right)
\right] \rho_0 ( r , \theta , r' , \theta' ; t ) .
\label{eq:MasterEq-ABring}
\end{align}
\end{widetext}
Here, the first, second, and third terms on the right-hand side correspond to the Liouvillian including the fluctuation term, and the dissipation terms in the $r$ and $\theta$ directions, respectively.

Reducing the radial degree of freedom, as shown in Eqs.~\eqref{eq:defReducedDensityOperator} and \eqref{eq:defReducedDensityOperator2}, leads to the following equation for the reduced density matrix $R_0(\theta, \theta'; t)$:
\begin{align}
&~ \frac{\partial}{\partial t} R_0 ( \theta , \theta' ; t ) \nonumber \\
&~~~~~~ = \left\{ - \frac{i}{\hbar} \hat{H}_S^\times - \frac{\eta r_0^2}{\beta \hbar^2} ( 1 - \cos ( \theta - \theta' ) )
\right\} R_0 ( \theta , \theta' ; t ) \nonumber \\
&
~~~~~ + \frac{\eta}{m_S} ( 1 - \cos ( \theta - \theta' ) ) R_0 ( \theta , \theta' ; t ) \nonumber \\
&
~~~~ - \frac{\eta}{2 m_S} \sin ( \theta - \theta' )
\left( \frac{\partial}{\partial \theta} - \frac{\partial}{\partial \theta'}
+ \frac{2 q r_0 A_\theta}{i \hbar} \right) R_0 ( \theta , \theta' ; t ) ,
\label{eq:MasterEq-ABring2}
\end{align}
where the second term originates from the dissipation in the radial direction, highlighting the coupling between radial and angular degrees of freedom. Since the fluctuation term and the second term share the same functional form, they can be unified by introducing an effective inverse temperature, $\beta' = \beta / \left(1 - \beta \hbar^2 / (2 I_S)\right)$, into the fluctuation term. Applying the DWT-PBC to Eq.~\eqref{eq:MasterEq-ABring2} yields the U(1)-QFPE presented in Eq.~\eqref{eqn:FP_RISB}.

\section{Caldeira--Legget model for the AB ring} \label{CL_ABring}
The Hamiltonian of the CL model\cite{CALDEIRA1983587} is presented in Eq. \eqref{eqn:C-L}
We consider the Drude SDF is then expressed as 
\begin{eqnarray}
 J(\omega) = \frac{\hbar \eta}{\pi}\frac{\gamma^2\omega}{\gamma^2+\omega^2}.
\label{JDrude2}
\end{eqnarray}
With a straightforward extension of the quantum hierarchical Fokker-Planck equation (QHFPE) for the Hamiltonian Eq. \eqref{eqn:C-L}, we obtain the CL U(1)-QHFPE expressed as\cite{T15JCP}
\begin{align}
\frac{\partial{W}_{\rm CL}^{(\bm n)} ( p , \theta )}{\partial t}
&= - \left[ \frac{p - q r_0 A}{I_S} \frac{\partial}{\partial \theta}
+ \sum_{k = 0}^K n_k \nu_k \right] W_{\rm CL}^{(\bm n)} ( p , \theta ) \nonumber \\
& 
+ \hat{\Phi} \sum_{k = 0}^K W_{\rm CL}^{(\bm n + \bm e ^k)} ( p , \theta)  \nonumber \\
	&+ \sum_{k = 0}^K n_k \nu_k \hat{\Theta}_k^{\rm CL} W_{\rm CL}^{(\bm n - \bm e ^k)} ( p , \theta) ,
\label{heom_wig2}
\end{align}
where $\hat{\Phi} =  r_0 {\partial} / {\partial p}$,
\begin{align}
\hat{\Theta}_0^{\rm CL} \equiv \frac{\eta r_0}{I_S}
\left\{ ( p - q r_0 A ) + \frac{I_S}{\beta} 
\left( 1 + \sum_{j = 1}^K \frac{2 \bar{\eta}_j \gamma^2}{\gamma^2 - \nu_j^2} \right)
\frac{\partial}{\partial p} \right\} ,
\end{align}
and
\begin{equation}
\hat{\Theta}_k^{\rm CL} \equiv 
- \frac{2 \eta \bar{\eta}_k \gamma^2}{\beta ( \gamma^2 - \nu_k^2 )} \ddp
\quad ( k = 1 , \cdots , K ) .
\end{equation}
The above equations are then truncated by using the modified ''terminators'' expressed in the Wigner representation.\cite{T15JCP}

For large $\gamma$ we have $J(\omega) = \hbar \eta \omega/\pi$.  With the high temperature bath $\beta \hbar \omega_0 \ll 1$, where $\omega_0$ is the characteristic frequency of the system, we have
\begin{eqnarray}
\frac{\partial W_{\rm CL} (p,\theta)}{\partial t} &&=  - 
\frac{p - q r_0 A}{I_{S}} \frac{\partial}{\partial \theta} W_{\rm CL}(p,\theta ) \nonumber \\
&&  + \frac{\eta}{m_S} \frac{\partial}{\partial p} \left[ ( p - q r_0 A ) + \frac{I_{S}}{\beta} \frac{\partial}{\partial p} \right] W_{\rm CL} ( p , \theta ) . \nonumber \\
\label{eqn:CLFPE}
\end{eqnarray}
It is important to notice that because the Hamiltonian Eq.\eqref{eqn:C-L} does not have rotational symmetry, we cannot employ the DWT-PBC presented in Eq. \eqref{eqn:DisWigTran}.  To impose periodic boundary conditions,  we thus set $W_{\rm CL}(p, \theta) = W_{\rm CL}(p, \theta + 2\pi)$.  
The above equation should be semi-classical, but $\hbar$ does not appear anywhere. Consequently, the description of the CL U(1)-QFPE with the periodic boundary condition is by nature classical.

\bibliography{tanimura_publist,AB,references}
\end{document}